\documentclass[a4paper,fleqn,usenatbib]{mnras}

\usepackage{newtxtext,newtxmath,verbatim}

\mathchardef\mhyphen="2D 

\newcommand{\acknowledgments}[1]{\begin{small}\section*{Acknowledgments}\end{small}{\noindent #1}\vspace{5pt}}

\newcommand{\Dt}[1]{\frac{d #1}{dt}}
\newcommand{\initvalupper}[1]{#1^{0}}
\newcommand{\initvallower}[1]{#1_{0}}
\newcommand{\driftvel}{{\bf w}_{s}}
\newcommand{\driftvelmag}{w_{s}}
\newcommand{\driftvelhat}{\hat{{\bf w}}_{s}}
\newcommand{\dustvel}{{\bf v}_{d}}
\newcommand{\gasvel}{{\bf u}_{g}}
\newcommand{\gasden}{\rho_{g}}
\newcommand{\dustden}{\rho_{d}}

\newcommand{\ts}{t_{s}}
\newcommand{\grainsize}{\epsilon_{d}}
\newcommand{\cs}{c_{s}}
\newcommand{\internaldensity}{\bar{\rho}_{d}^{\,i}}
\newcommand{\tL}{t_{L}}

\newcommand{\B}{{\bf B}}
\newcommand{\Bhat}{\hat{\bf B}}
\newcommand{\acc}{{\bf a}}
\newcommand{\Lbox}{L_{\rm box}}
\newcommand{\Alf}{Alfv\'en}
\newcommand{\mpanel}[1]{\includegraphics[width=0.24\textwidth]{#1.jpeg}}
\newcommand{\grainsizedl}{\bar{\epsilon}_{d}}
\newcommand{\acceldl}{\bar{\rm a}}
\newcommand{\grainchargedl}{\bar{\phi}_{d}}

\usepackage[T1]{fontenc}
\usepackage{ae,aecompl}

\usepackage{graphicx}	
\usepackage{amsmath}	
\usepackage{amssymb}	

\title[Magnetic RDIs]{Nonlinear Evolution of the  Resonant Drag Instability in Magnetized Gas}

\author[Seligman, Hopkins, \& Squire]{
Darryl Seligman,$^{1}$\thanks{E-mail: darryl.seligman@yale.edu}
Philip F. Hopkins,$^{2,3}$
and Jonathan Squire$^{2,3,4}$
\\
$^{1}$Department of Astronomy, Yale University, 52 Hillhouse Ave., New Haven CT 06511, USA \\
$^{2}$TAPIR, Mailcode 350-17, California Institute of Technology, Pasadena, CA 91125, USA \\
$^{3}$Walter Burke Institute for Theoretical Physics, Pasadena, CA 91125, USA \\
$^{4}$Department of Physics, University of Otago, P.O.~Box 56, Dunedin 9054, New Zealand \\
}

\date{Accepted XXX. Received YYY; in original form ZZZ}

\pubyear{2015}

\begin{document}
\label{firstpage}
\pagerange{\pageref{firstpage}--\pageref{lastpage}}
\maketitle

\begin{abstract}
We investigate, for the first time, the nonlinear evolution of the magnetized ``resonant drag instabilities'' (RDIs). We explore magnetohydrodynamic (MHD) simulations of gas mixed with (uniform) dust grains subject to Lorentz and drag forces, using the {\small GIZMO} code. The magnetized RDIs exhibit fundamentally different behaviour than the purely acoustic RDIs. The dust organizes into coherent structures and the system exhibits strong dust-gas separation. In the linear and early nonlinear regime, the growth rates agree with linear theory and the dust self-organizes into two-dimensional planes or ``sheets.'' Eventually the gas develops fully nonlinear, saturated \Alf{ic} and compressible fast-mode turbulence, which fills the under-dense regions with a small amount of dust, and drives a dynamo which saturates at equipartition of kinetic and magnetic energy. The dust density fluctuations exhibit significant non-Gaussianity, and the power spectrum is strongly weighted towards the largest (box-scale) modes. The saturation level can be understood via quasi-linear theory, as the forcing and energy input via the instabilities becomes comparable to saturated  tension forces and dissipation in turbulence. The magnetized simulation presented here is just one case; it is likely that the magnetic RDIs can take many forms in different parts of parameter space.
\end{abstract}

\begin{keywords}
Instabilities--turbulence-- ISM: kinematics and dynamics --star formation: general--galaxies: formation--
cosmology: theory-- planets and satellites: formation-- accretion, accretion disks
\end{keywords}



\vspace{-0.5cm}
\section{Introduction}

Most astrophysical fluids contain a spectrum of solid grains, commonly referred to as dust. This dust contains a large fraction of the metals in the Universe, and is prominent in the Interstellar Medium \citep{Ferriere2001}, protoplanetary disks \citep{Armitage2011}, and our Solar System \citep{Krueger2015}. Dust physics is also key to understanding extinction and reddening in radiative transfer, feedback and winds for both star formation and AGN, galactic-chemistry, stellar evolution, interstellar heating and cooling, and more. 

Dust-gas interactions have been extensively studied in planet formation. The meter barrier, or how micron size dust grains  evolve into kilometer sized planetesimals, has been a fundamental challenge for planet formation theory \citep{Goldreich1973,Chiang2010}. Once the planetesimals approach a millimeter in size, they are more likely to shatter upon collisions \citep{Blum2008}, and  aerodynamic drag initiates rapid migration into the  host star \citep{Adachi1976}. \citet{Youdin2005} proposed the so-called ``streaming instability'' as a promising solution to the meter sized barrier, where the aerodynamic interaction between dust and gas causes an instability that has a growth rate  quicker than the  migration timescale, clumping grains and helping their coagulation into planetesimals \citep{Johansen2007Nature}. 

The physics of the streaming instability was generalized to a wide variety of other astrophyiscal systems by
\cite{Squire2018}, who demonstrated that dust grains streaming through a fluid are \textit{generically} unstable, if the dust streams faster than any fluid wave. The condition for the system to be unstable is, simply,  that the velocity of the dust grains projected along some direction matches the phase velocity of a linear fluid wave. These ``Resonant Drag Instabilities'' (RDIs) operate with almost any type of fluid oscillation; for instance sound waves, magnetosonic waves, epicyclic oscillations, and Brunt-V\"ais\"al\"a oscillations each create their own associated RDI. 

\citet{Hopkins2018b} presented a detailed linear analysis of the ``acoustic'' RDIs: RDIs in the simple case where the  gas mode is a pure-hydrodynamic sound wave, and the dust is uncharged. \citet{Squire2018b} explored RDIs of neutral grains relevant to planetesimal formation in proto-planetary disks, including the streaming instability, which is an RDI associated with the epicylic oscillations, and new instabilities including resonances with vertical settling, non-ideal MHD and buoyancy oscillations. \citet{Hopkins2018} further extended this by presenting a linear analysis of the case of charged dust in magnetized gas. They showed that gas obeying ideal MHD with charged dust grains (coupled to gas via generic gas drag and Lorentz forces) is \textit{always} unstable, at all wavelengths and for any non-zero gas to dust ratio, magnetic field strength, dust charge, and drift velocity. They identified several sub-families of magnetically driven instabilities, including the ``MHD-wave'' RDIs (resonance between dust advection and magnetosonic or \Alf\ waves), ``gyro'' RDIs (resonance between dust gyro motion and magnetosonic or \Alf\ waves), acoustic modes (akin to those in \citealt{Hopkins2018b}), ``pressure-free'' modes (which act on long wavelengths where magnetic pressure effects are weak), and ``cosmic ray-like'' modes (akin to resonant and non-resonant cosmic ray streaming instabilities; \citealp{Kulsrud1969,Bell:2004cj}). However, their analysis was ultimately limited to linear perturbation theory, but these instabilities cannot be linked to realistic physics or observations without understanding their nonlinear behavior. This requires numerical simulations.

In this paper, we present the first simulations of the nonlinear evolution of the magnetized RDI. This is part of a larger body of work that will elucidate the nonlinear evolution of the magnetized RDIs in different parts of parameter space. Our first focus here is a case study of a single initial condition, but we will show that this exhibits a rich, complicated nonlinear behavior with a variety of distinct competing modes present with similar growth rates.

\vspace{-0.5cm}
\section{Methods}

\begin{figure}
\begin{centering}
\includegraphics[width=0.84\columnwidth]{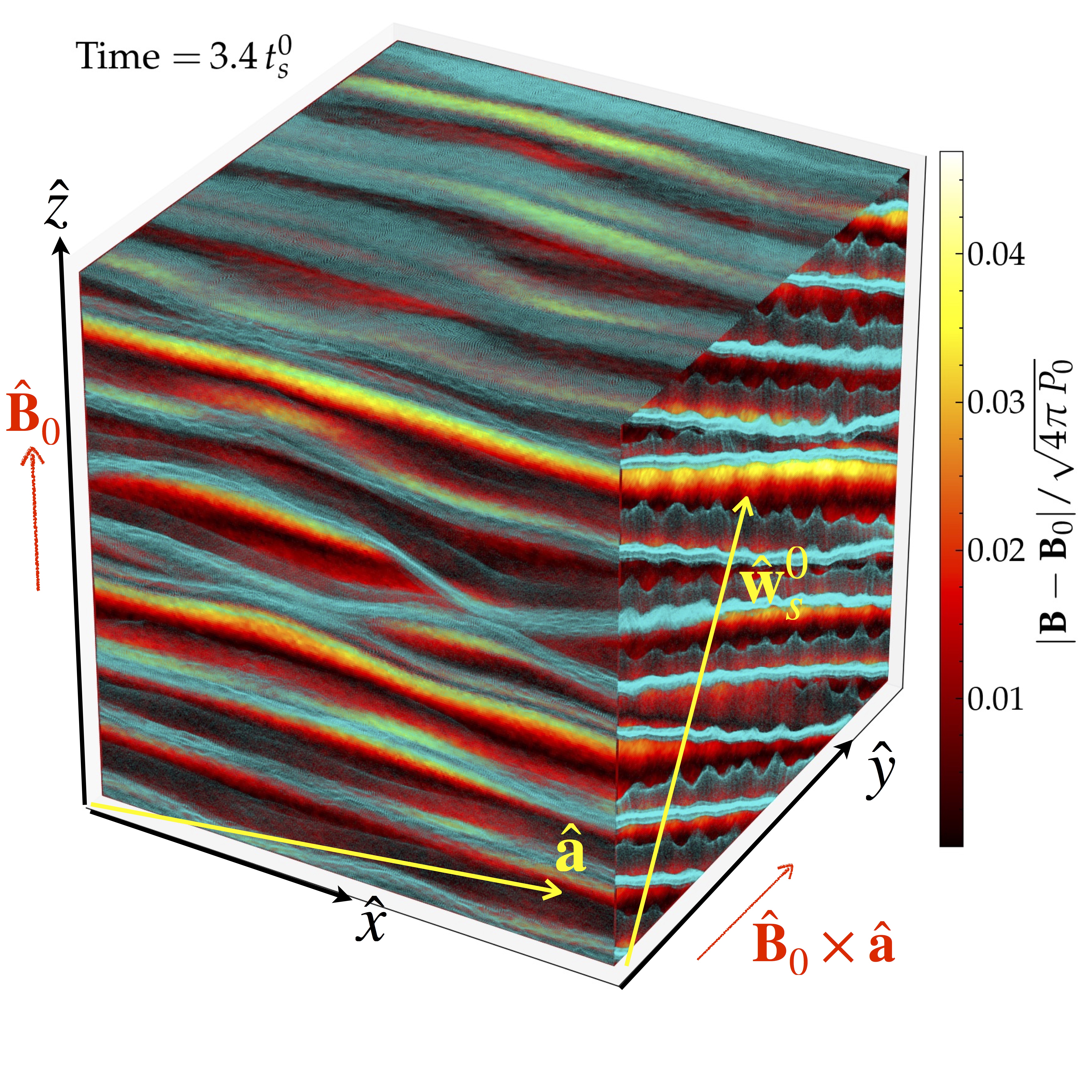}
\includegraphics[width=0.84\columnwidth]{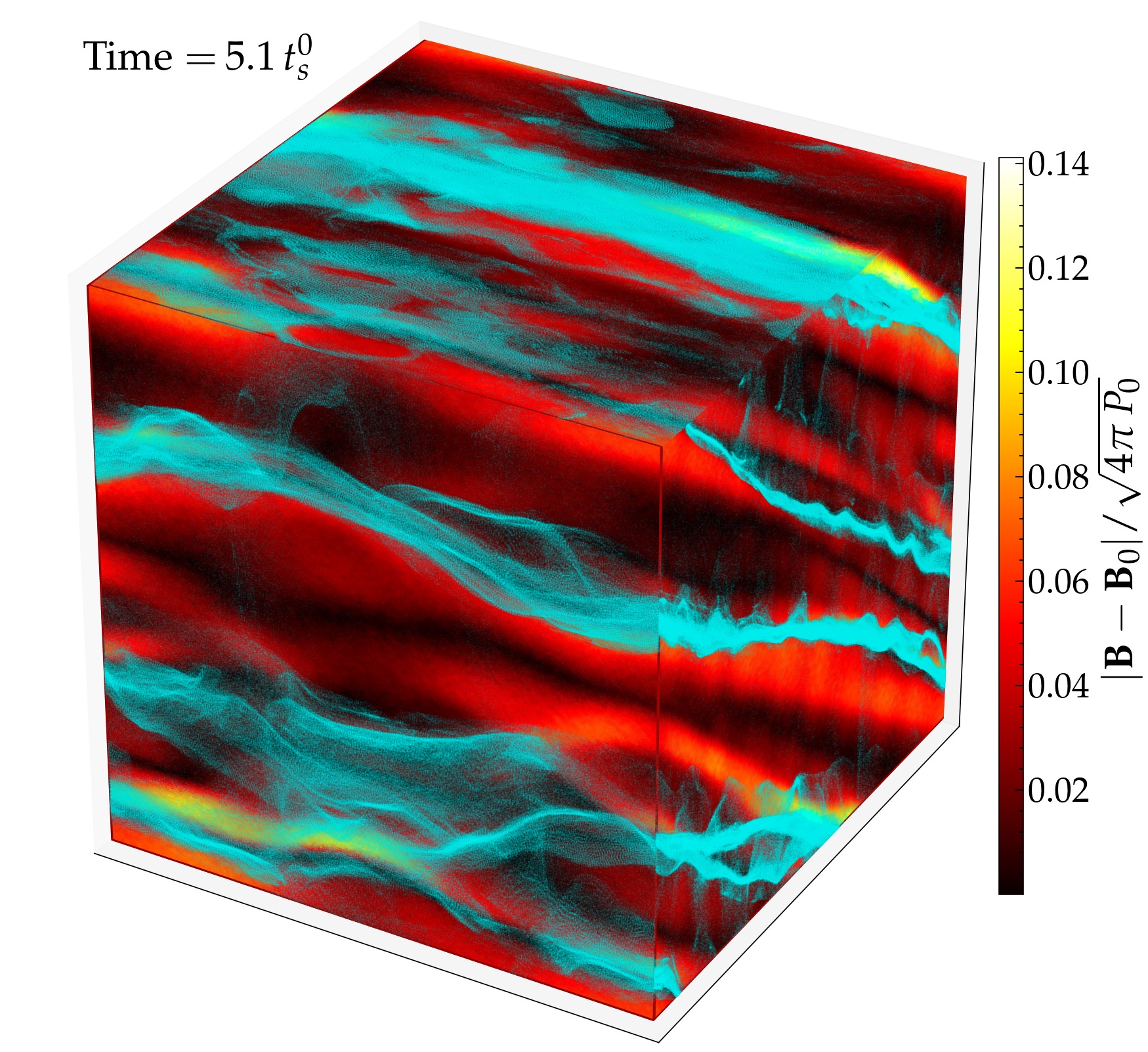}
\includegraphics[width=0.84\columnwidth]{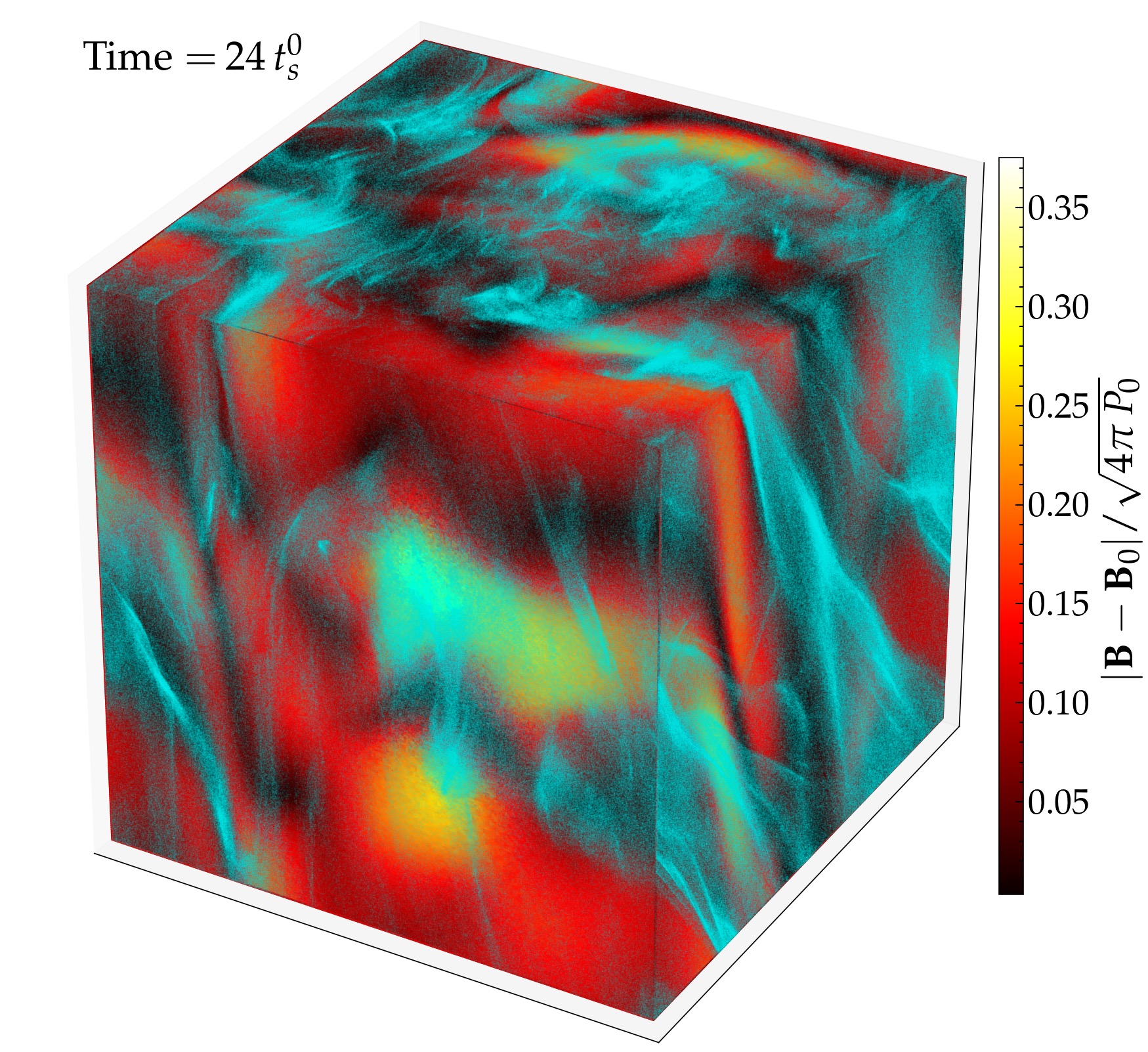}
\end{centering}\vspace{-0.2cm}
\caption{Three-dimensional visualization of the magnetic RDI. We show snapshots at three times (in units of the initial stopping/drag time) corresponding to linear ({\em top}), early nonlinear ({\em middle}), and saturated ({\em bottom}) regimes. Colors show the strength of magnetic field fluctuations; light-blue points show the locations of dust particles, on slices through each axis. Arrows on top panel indicate coordinates with initial field direction $\hat{\B}_{0}$, grain acceleration direction $\hat{\bf a}$, and equilibrium drift direction $\hat{\bf w}_{s}^{0}$. Dust self-organizes into ``sheets'' at the onset of instability, which persist through saturation. Gas turbulence grows through linear and early nonlinear phases and sustains itself in saturation, generating substructure in $\B$.}
\label{fig:3d_images}
\end{figure}

\begin{figure*}
\begin{centering}
\mpanel{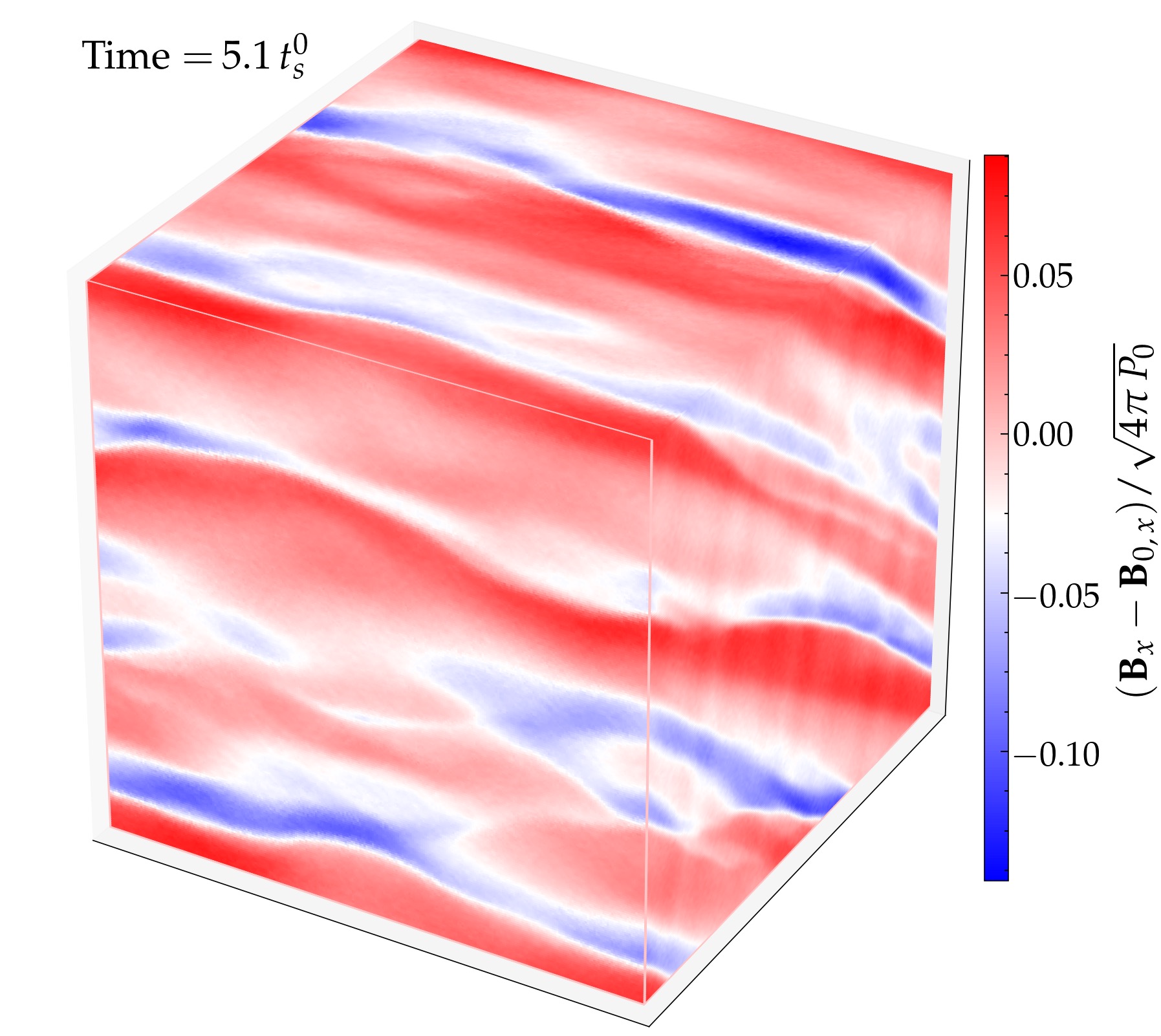}
\mpanel{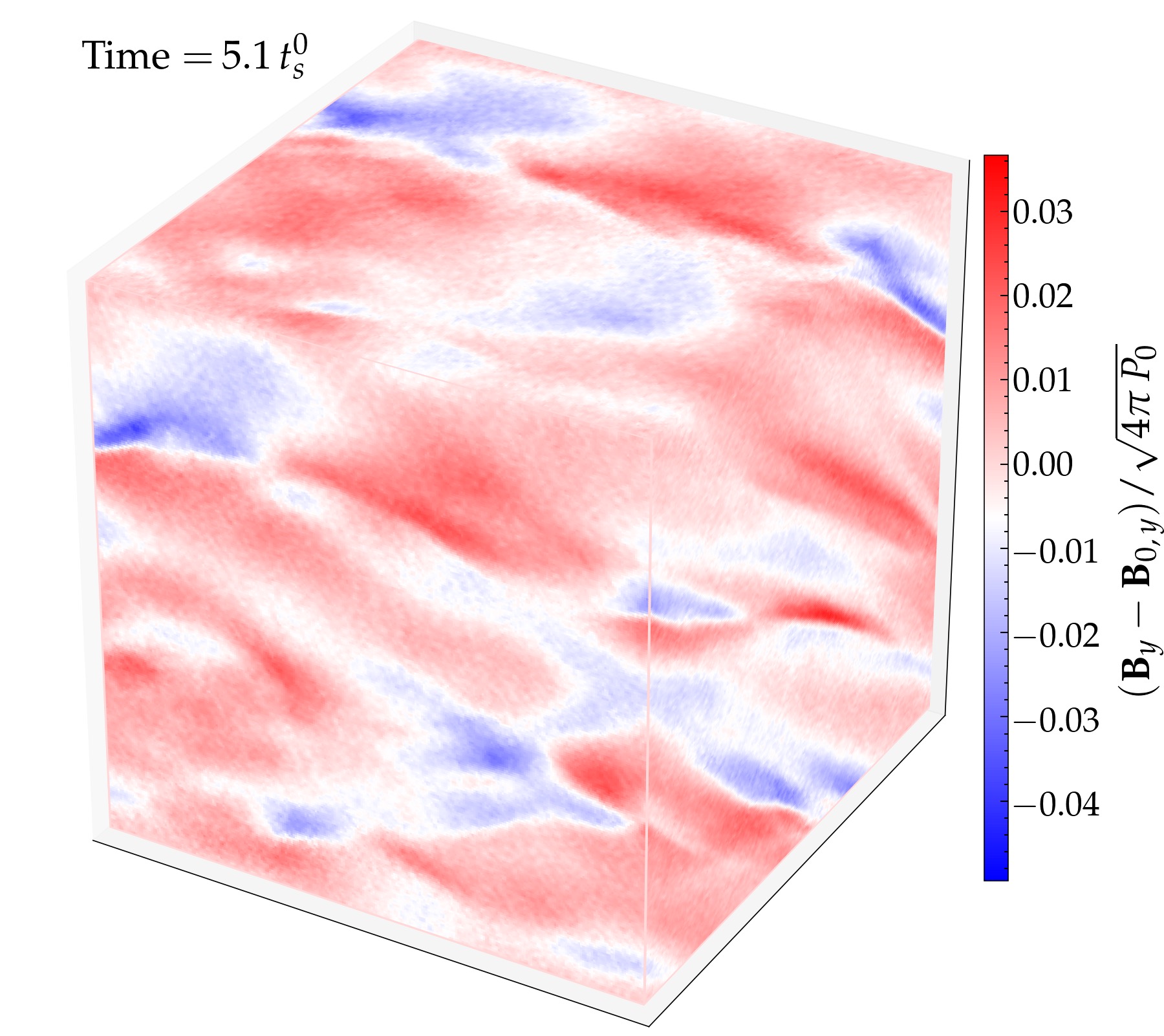}
\mpanel{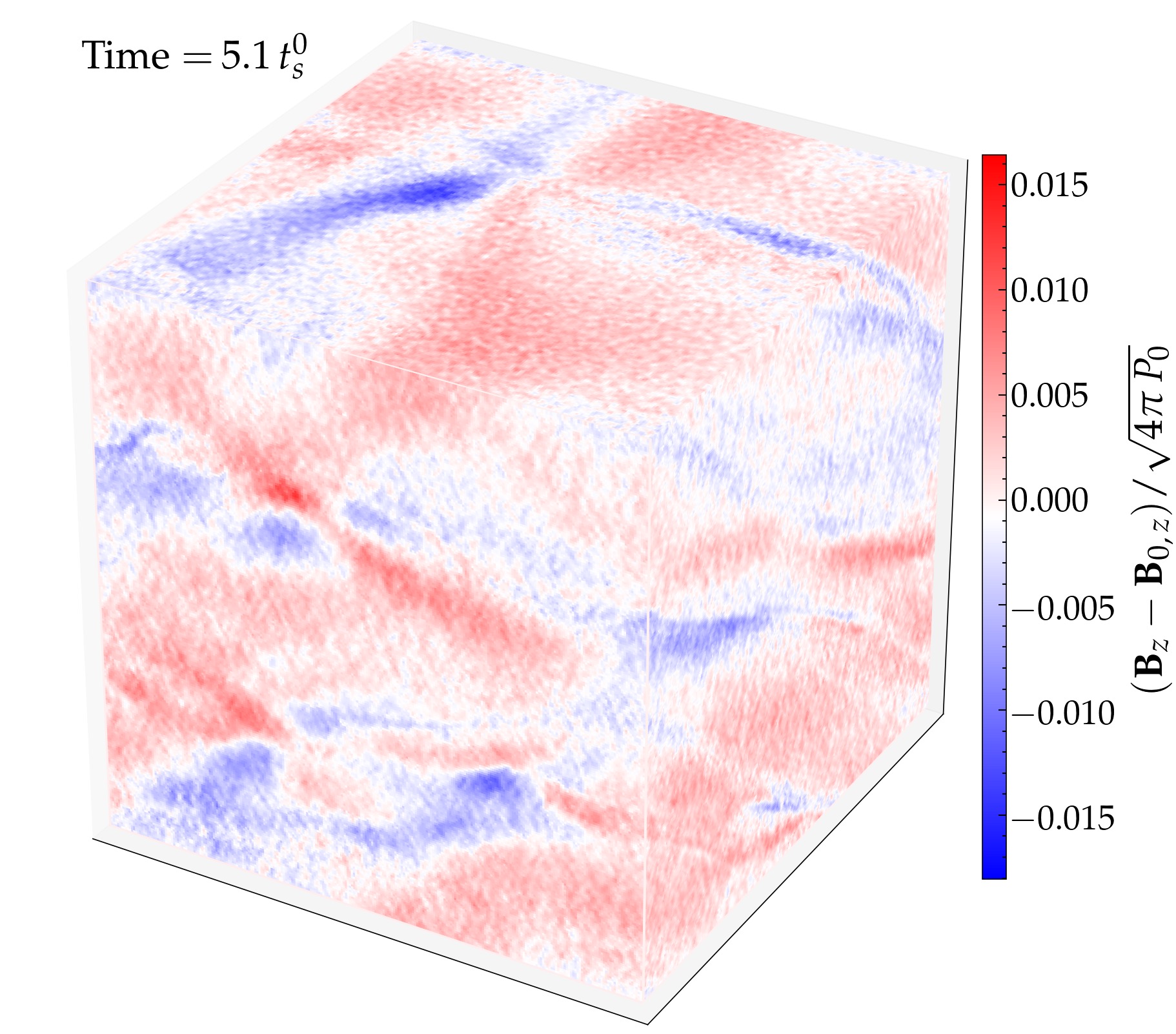} 
\includegraphics[width=0.24\textwidth]{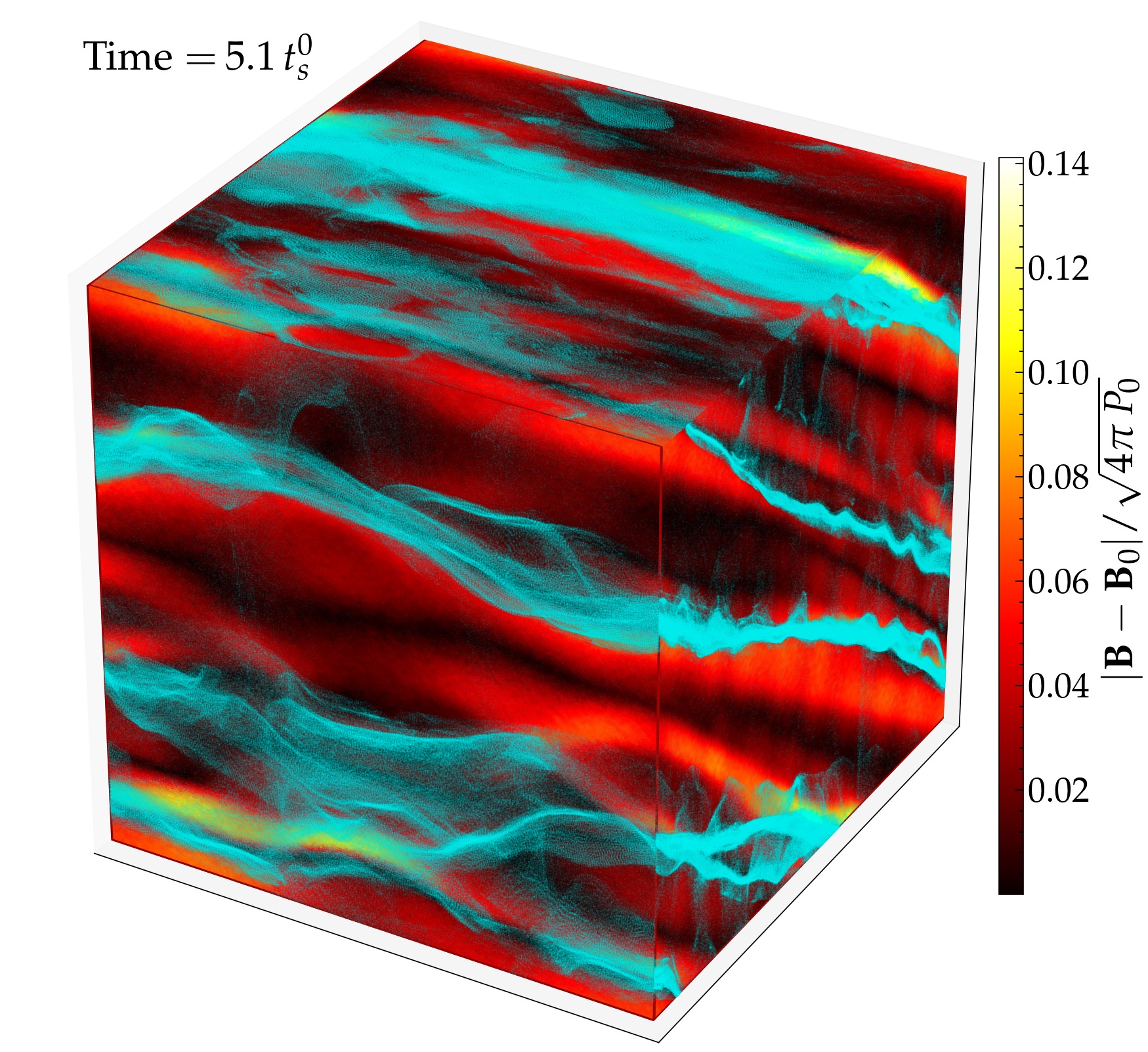}\\
\mpanel{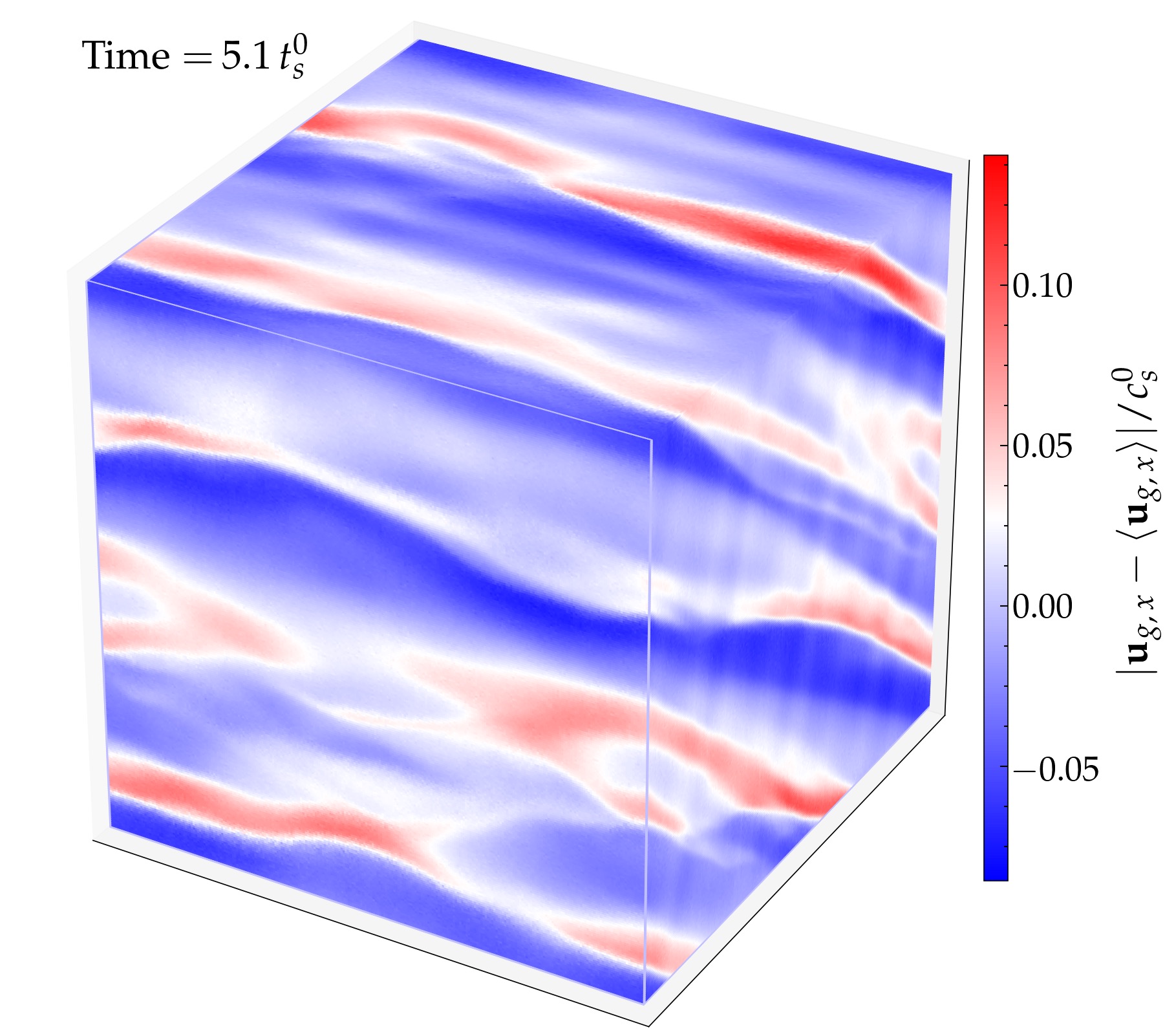}
\mpanel{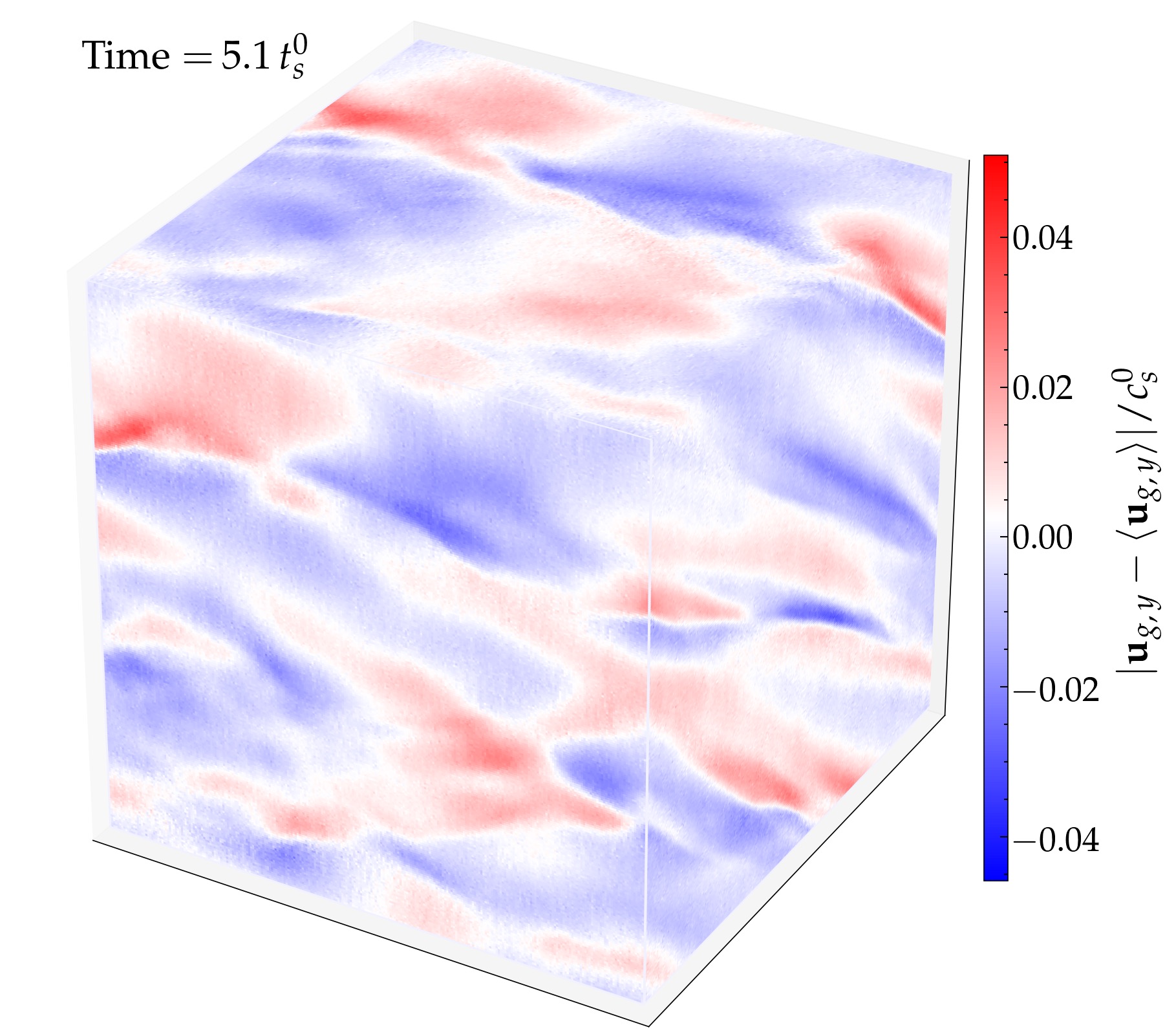}
\mpanel{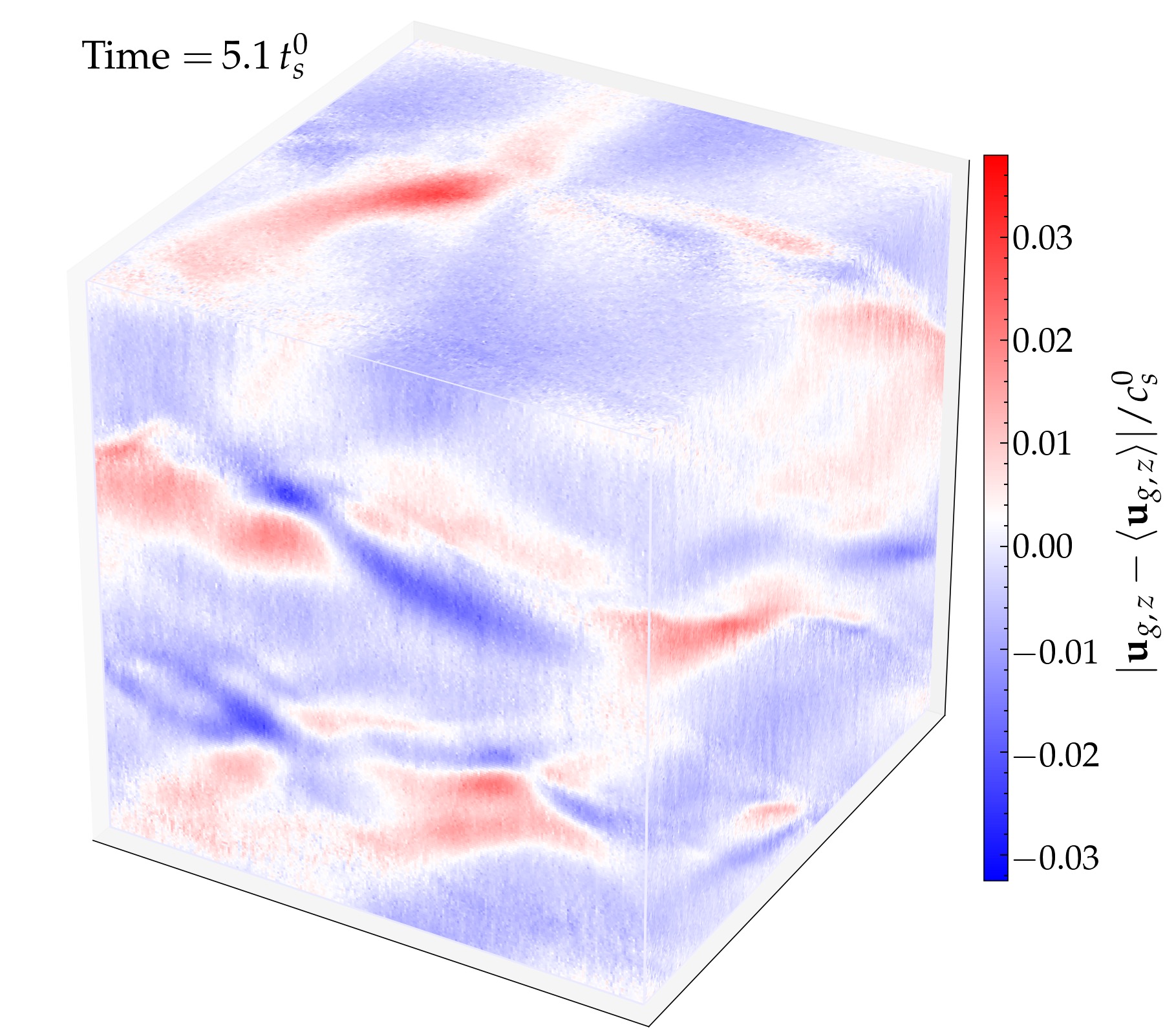} 
\mpanel{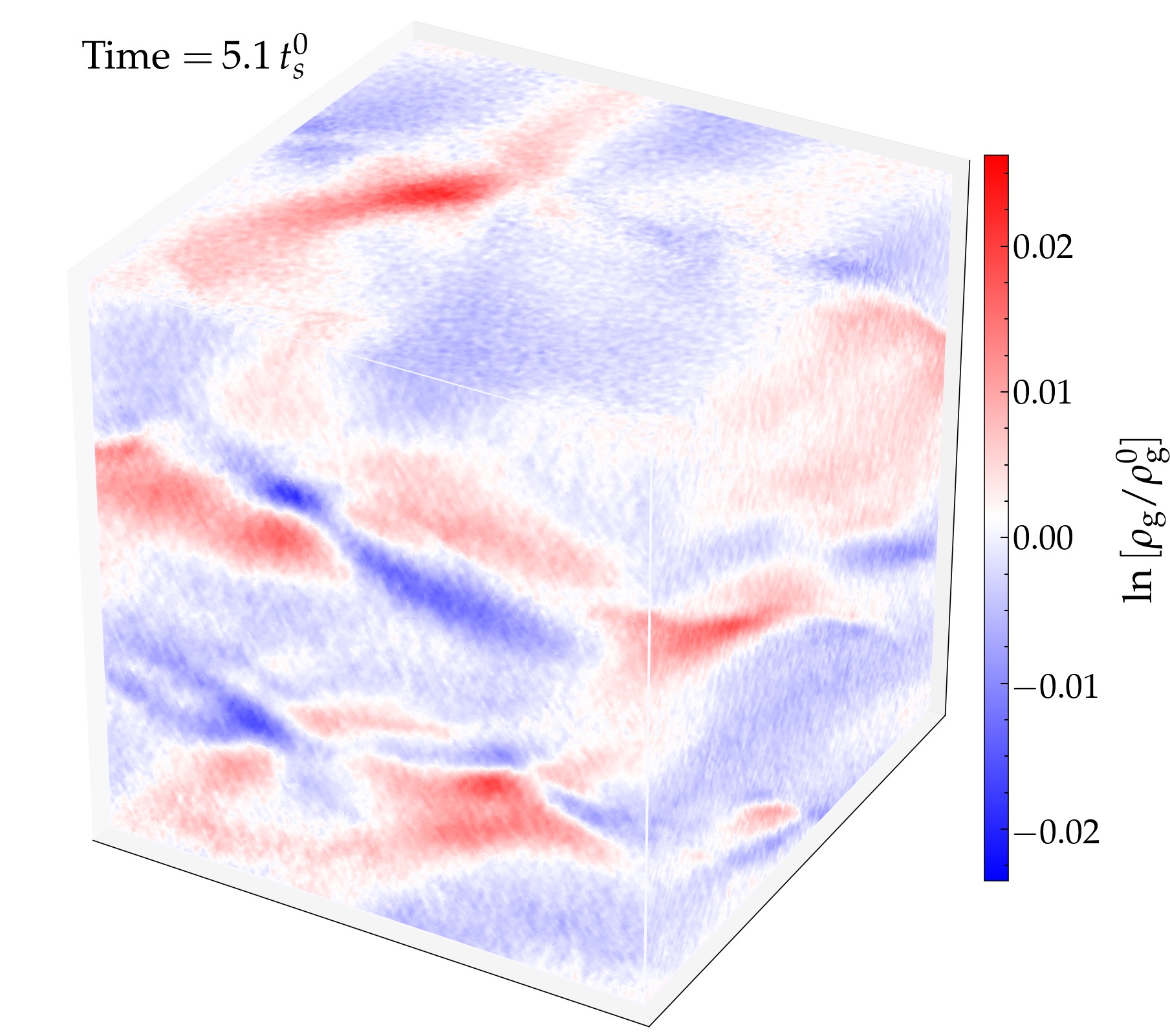}\\
\mpanel{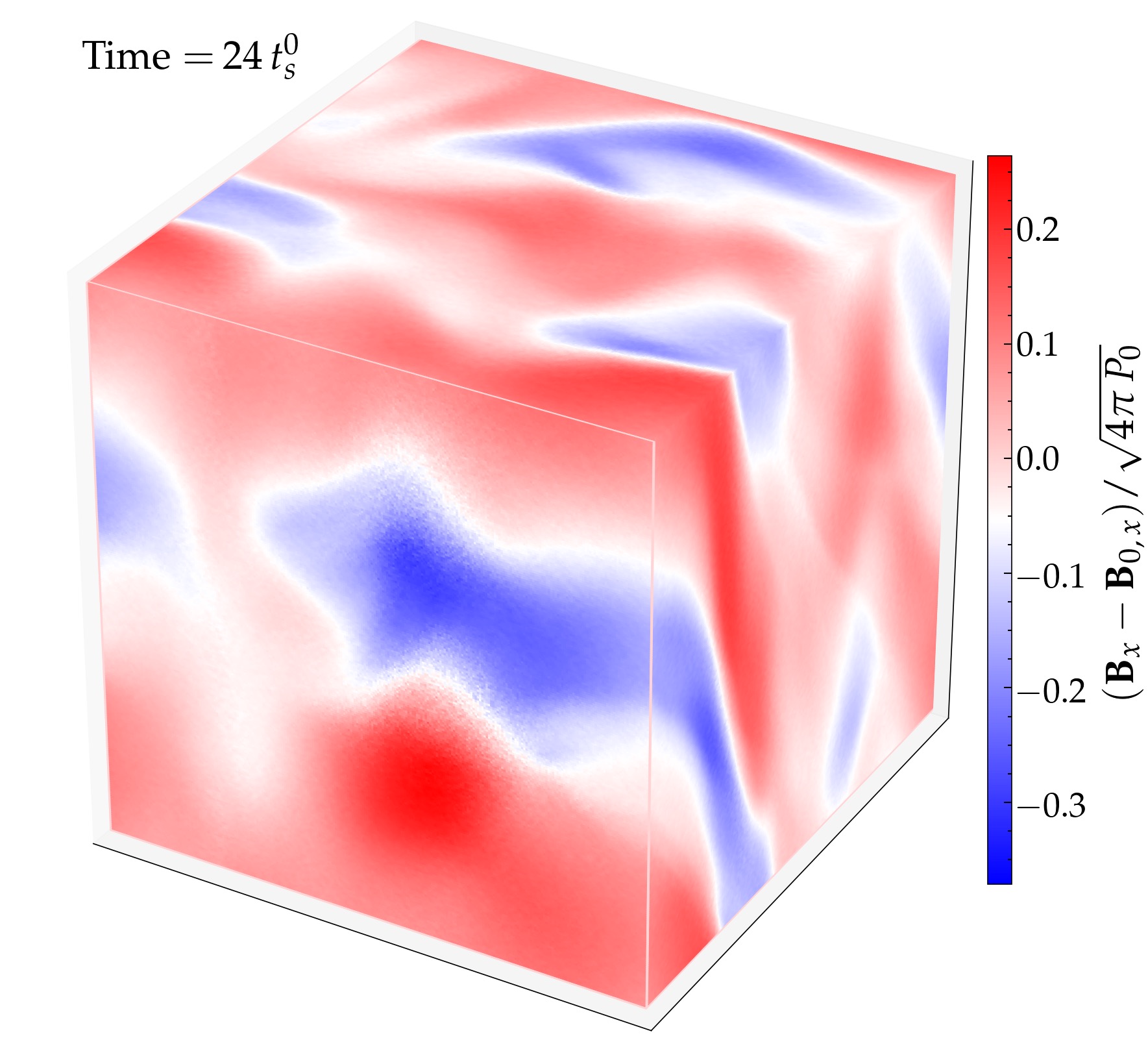}
\mpanel{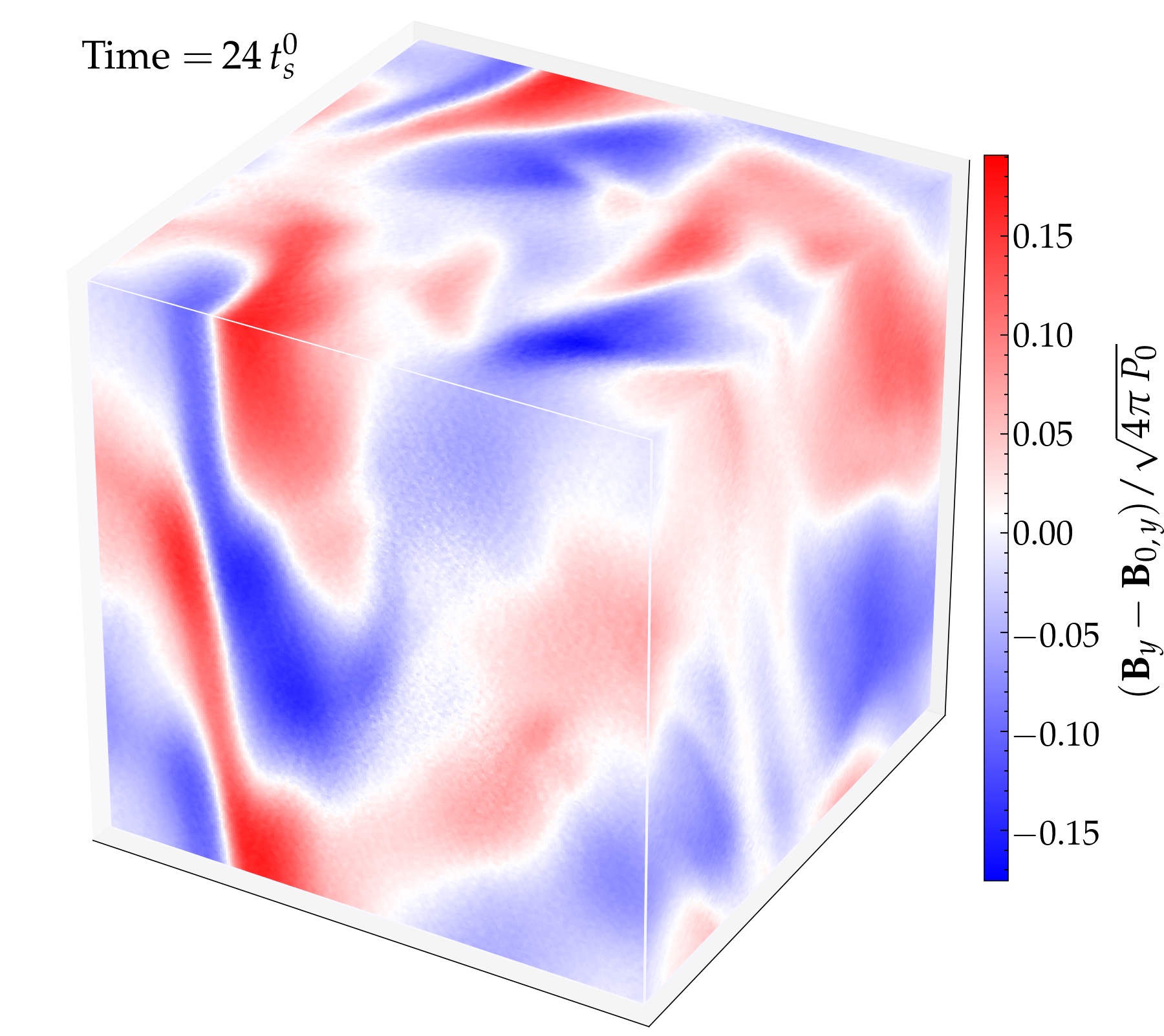}
\mpanel{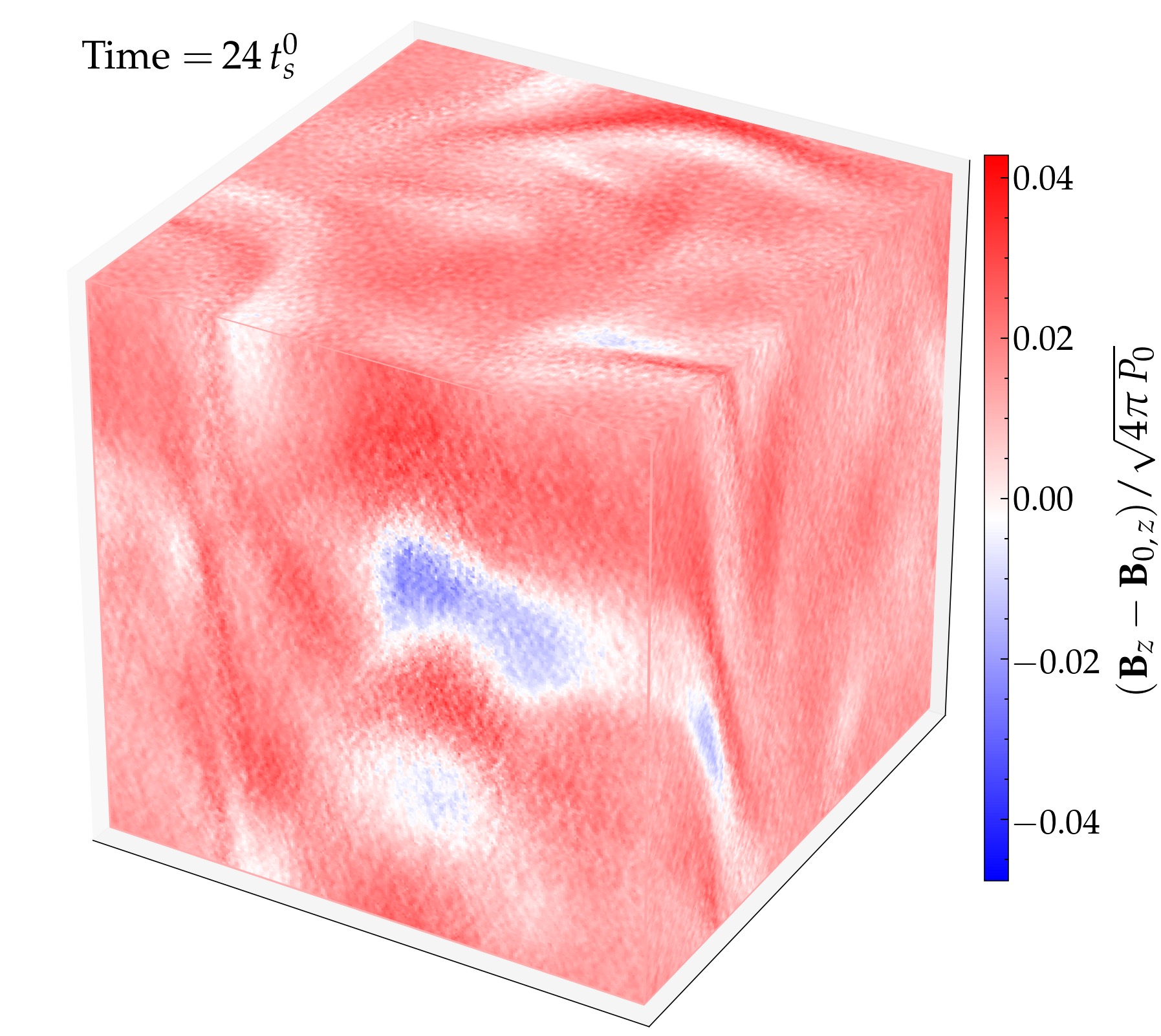} 
\includegraphics[width=0.24\textwidth]{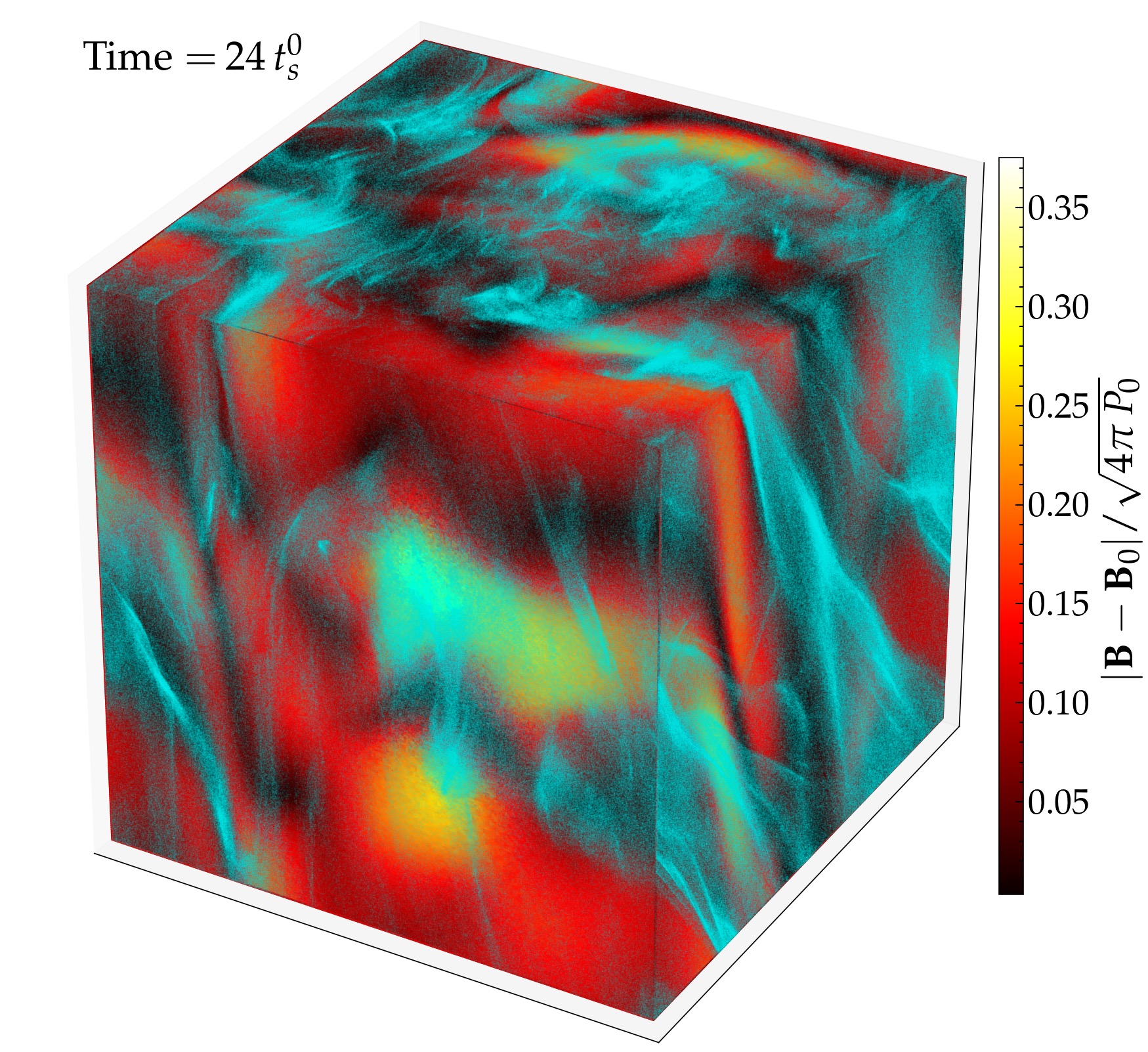}\\
\mpanel{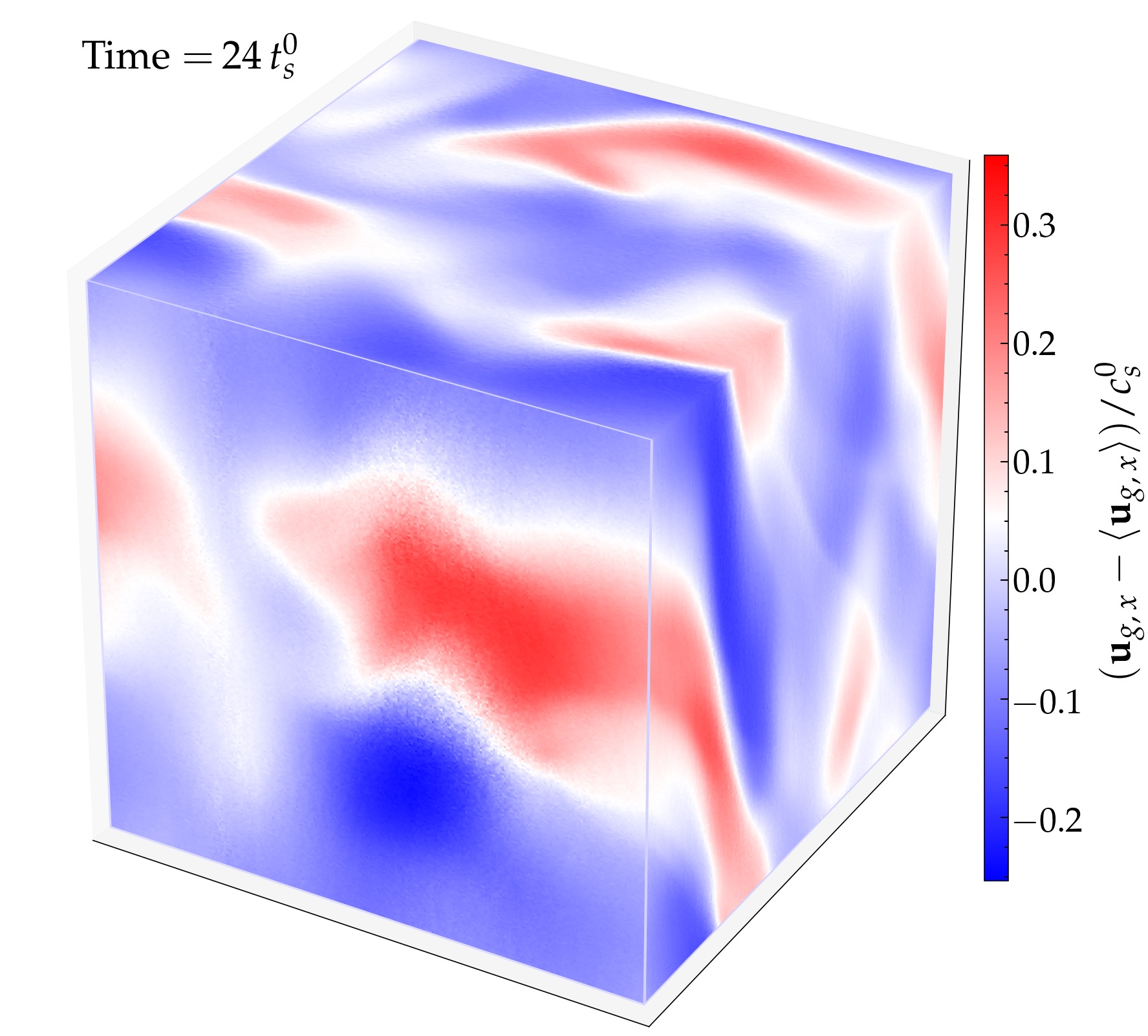}
\mpanel{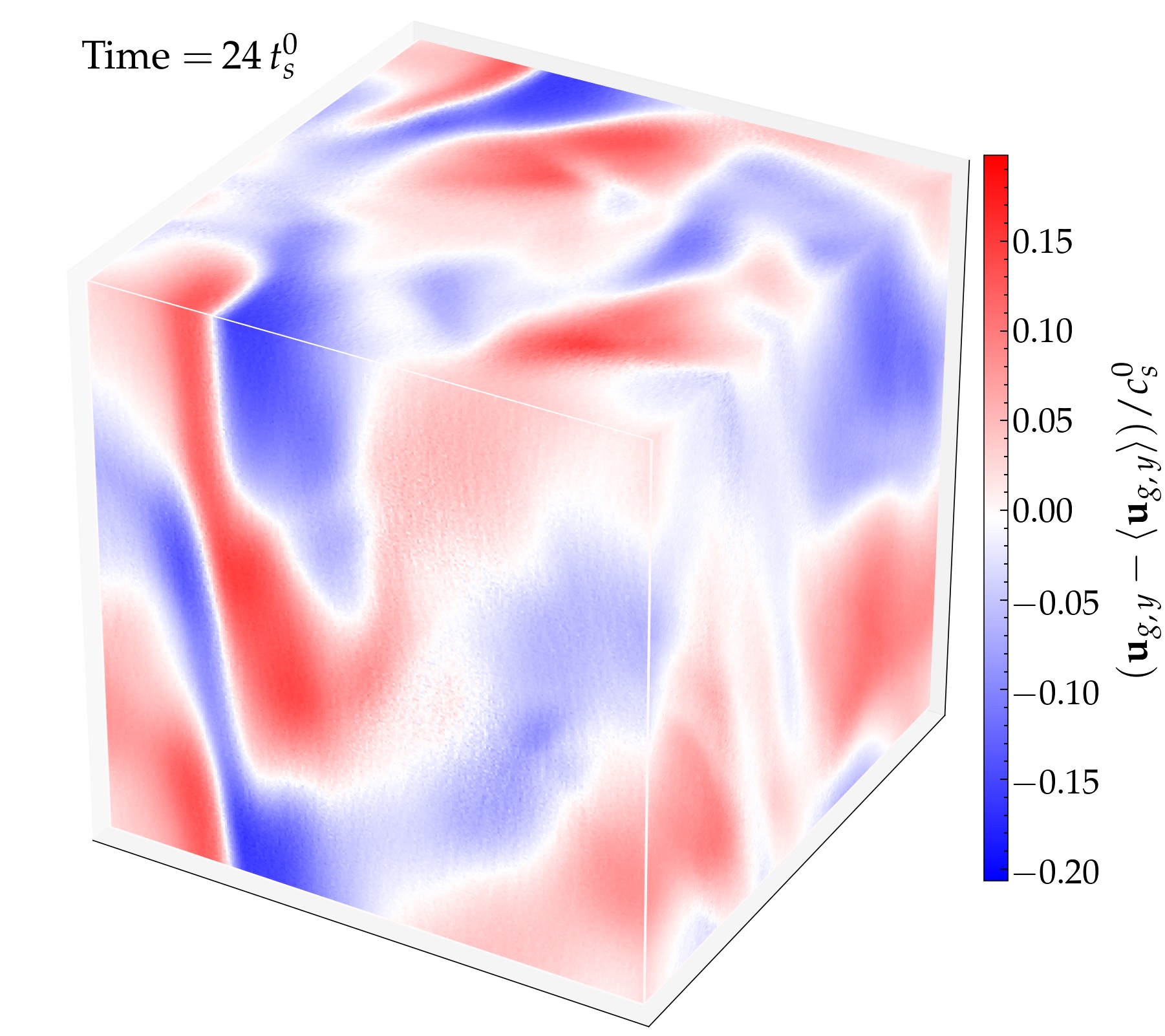}
\mpanel{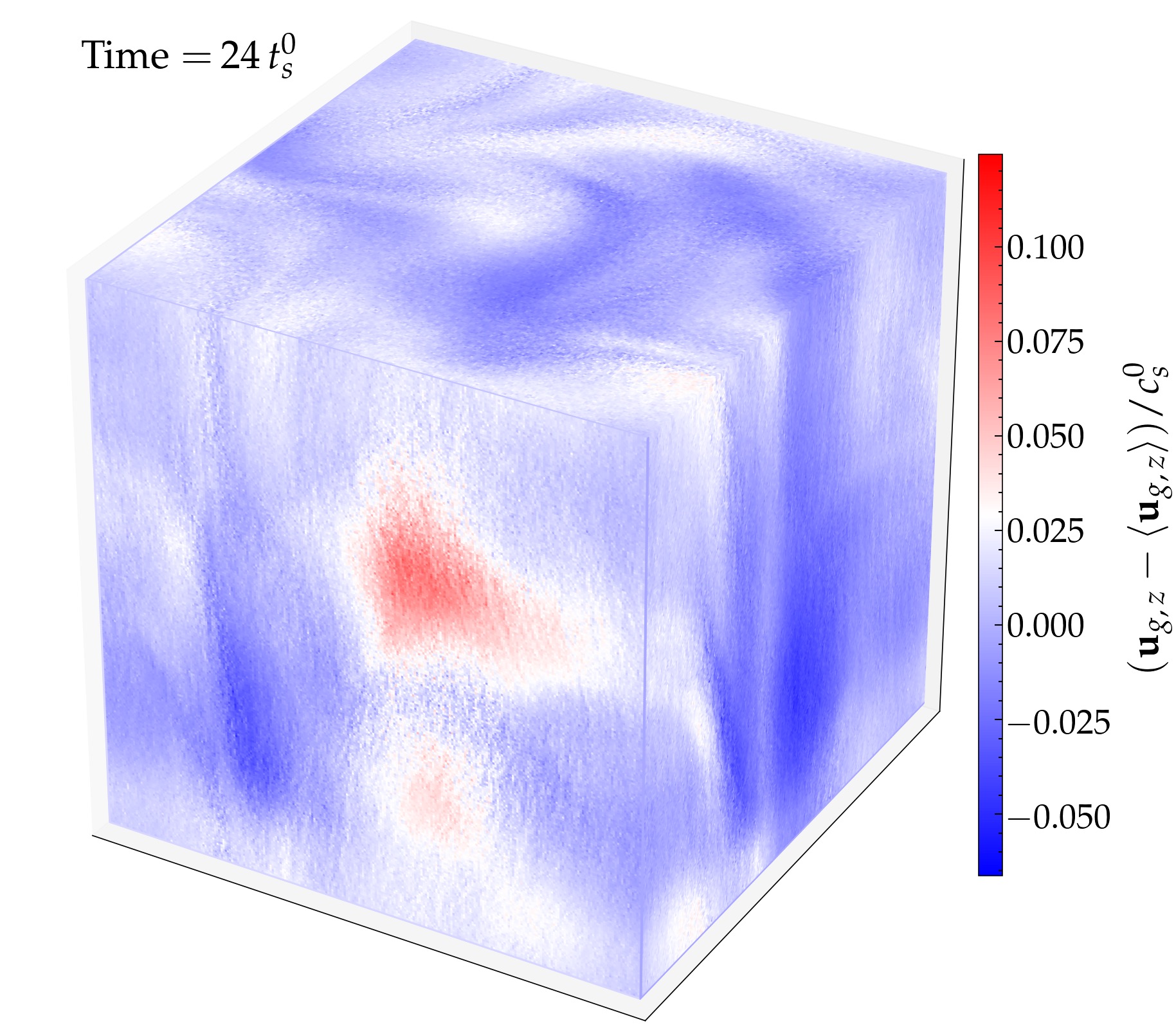} 
\mpanel{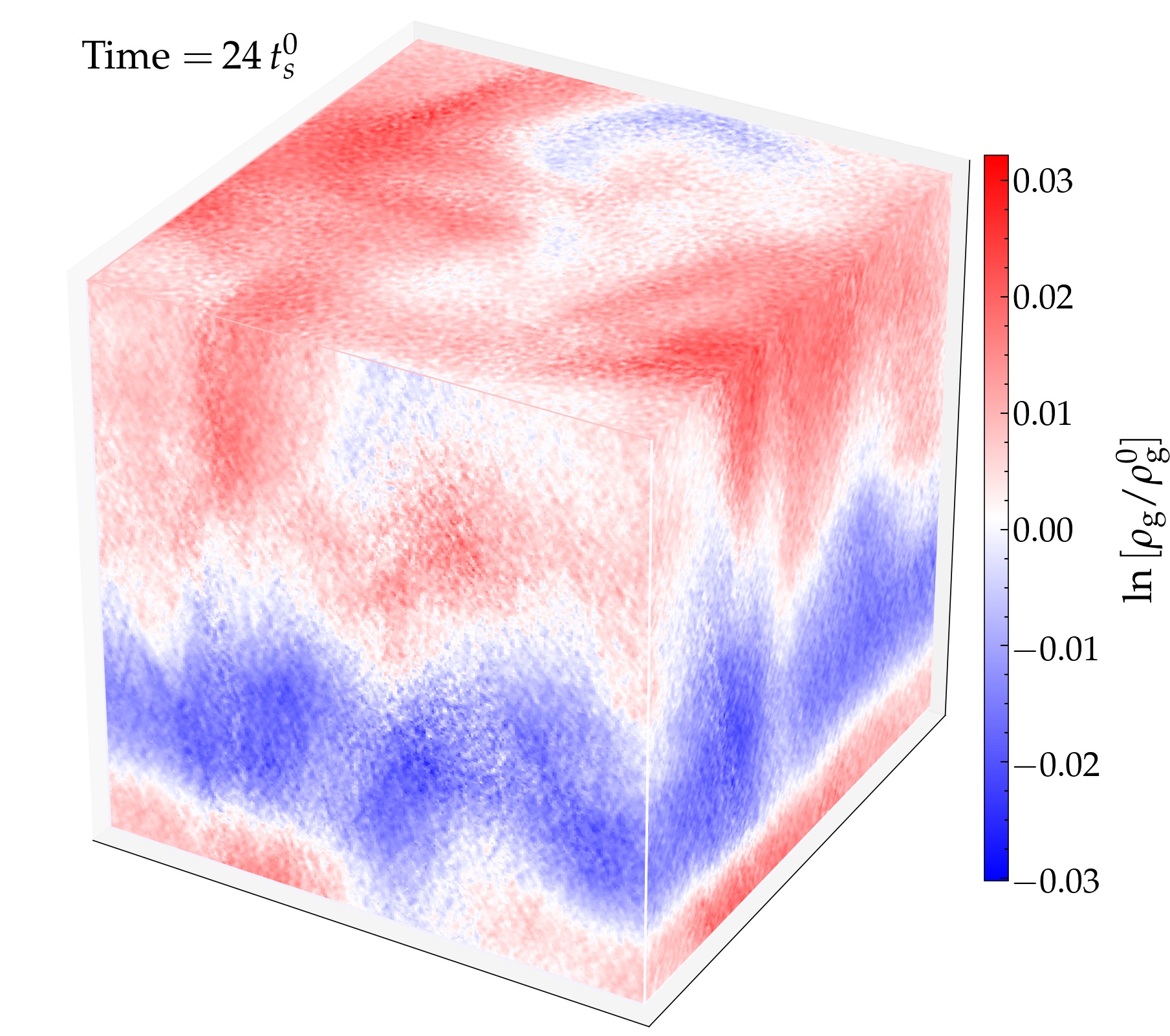}\\
\end{centering}\vspace{-0.2cm}
\caption{3D projections (as Figure \ref{fig:3d_images}) of the magnetic field ($\B$) and gas velocity ($\gasvel$) components, gas density $\gasden$, and the same dust density plot from Figure \ref{fig:3d_images}, at the ``early nonlinear'' time ({\em top}) and ``saturated'' time ({\em bottom}). The most prominent gas $\gasvel$/$\B$ fluctuations are clearly driven by the dust ``sheets'' as the dust ``slides'' along the sheet plane (while the sheet coherently drifts vertically along the field), dragging gas to ${\bf u}_{x} > 0$ within the sheet and bending the field lines. Each velocity component closely anti-correlates with the corresponding magnetic component as predicted by linear theory, with $(\B-\B_{0}) \approx -\hat{\bf k} \times ({\bf u} \times {\bf B}_{0}) / v_{p}$ (where $v_{p} = |\omega/k|$ is the phase velocity). Gas density fluctuations closely trace the parallel (compressible) $\B_{z}$ (and therefore ${\bf u}_{z}$) fluctuations, as in fast modes.}
\label{fig:2d_images}
\end{figure*}

\begin{figure}
\begin{centering}
\includegraphics[width=0.87\columnwidth]{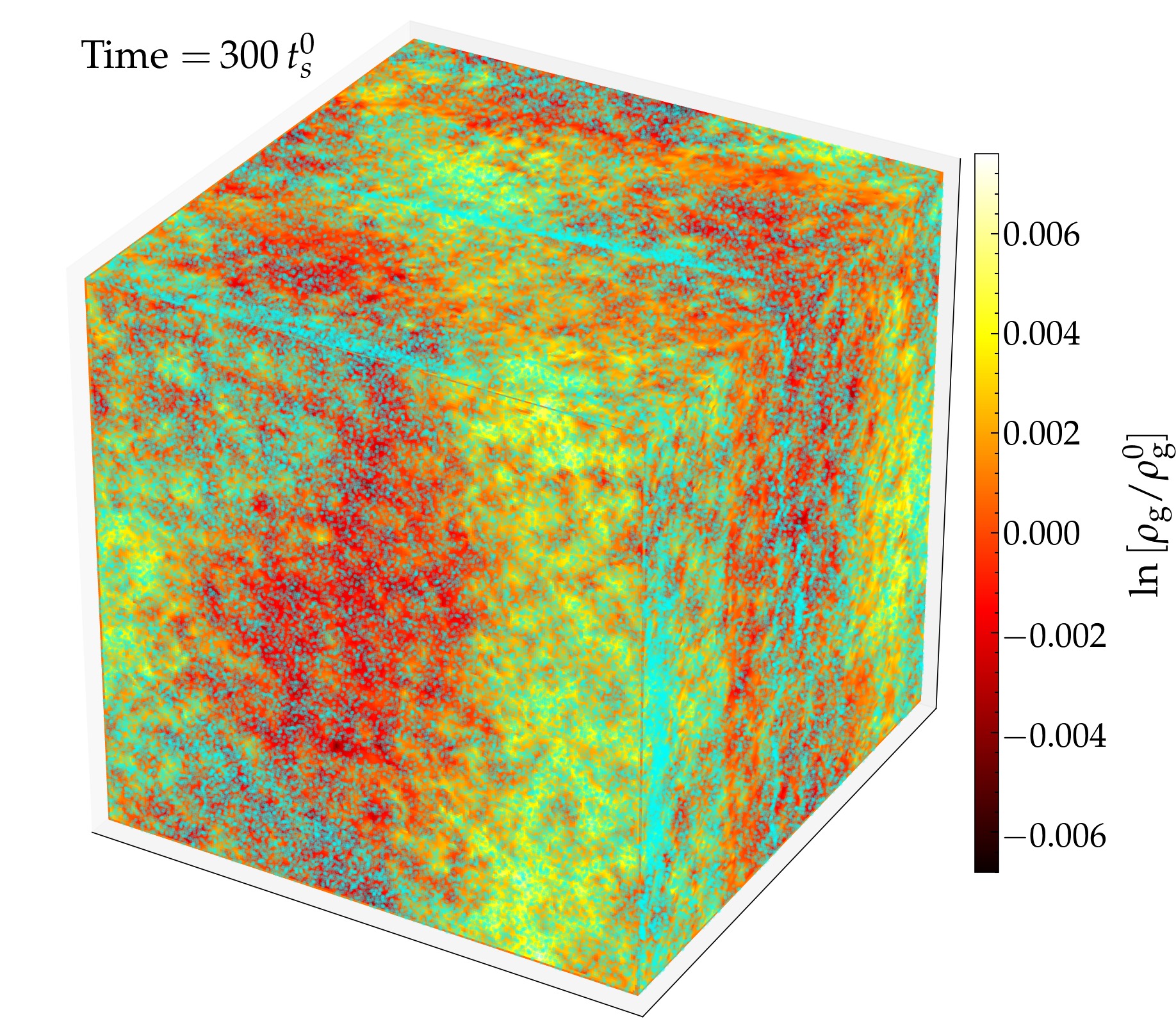}
\end{centering}\vspace{-0.2cm}
\caption{3D visualization as Fig.~\ref{fig:3d_images}, for a pure-hydro simulation (un-charged grains, $\B=0$). Color shows gas density. The equilibrium drift velocity is similar to Fig.~\ref{fig:3d_images}, as is the magnitude of the saturated turbulence in $\gasvel$, although the growth rate is slower (image is at the same number of growth times as Fig.~\ref{fig:3d_images}, {\em bottom}). The structure of the saturated dust structures (and earlier nonlinear structures) is entirely different.}
\label{fig:3d_images_hydro}
\end{figure}

\begin{figure}
\begin{centering}
\includegraphics[width=0.99\columnwidth]{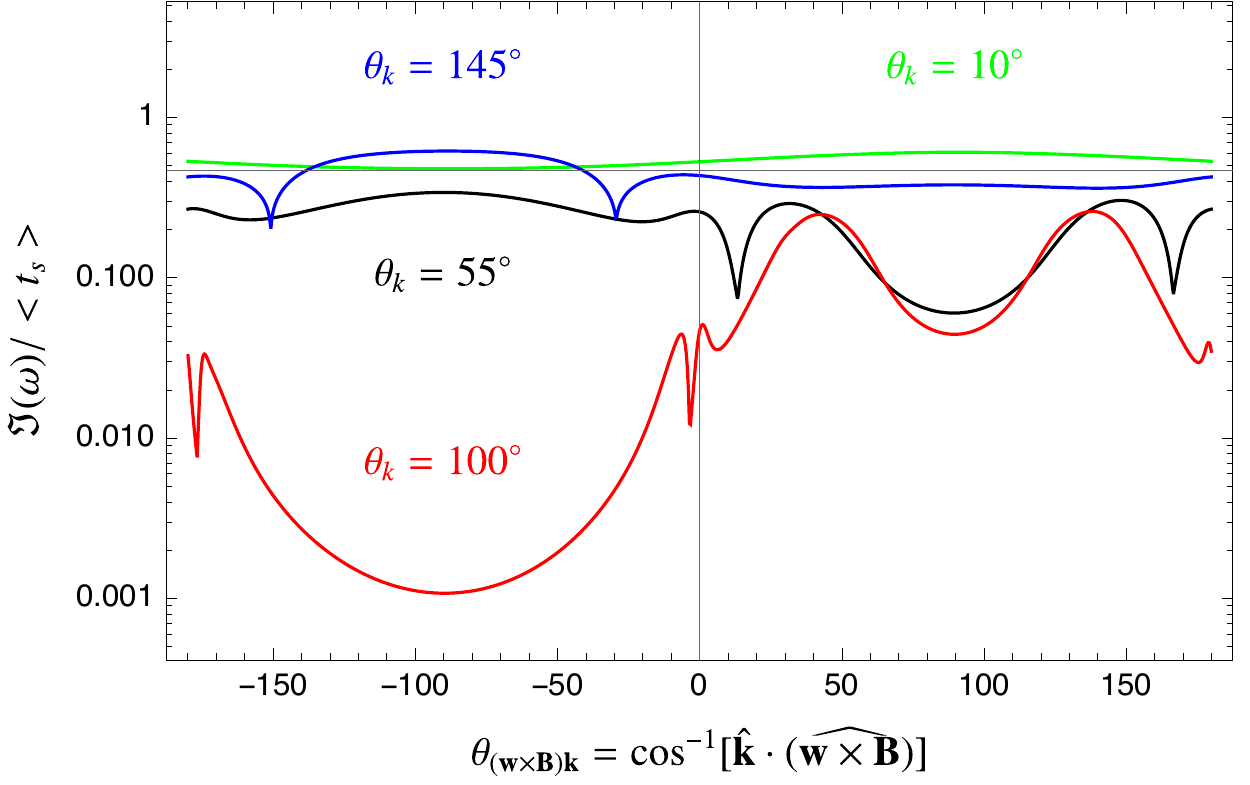}
\includegraphics[width=0.99\columnwidth]{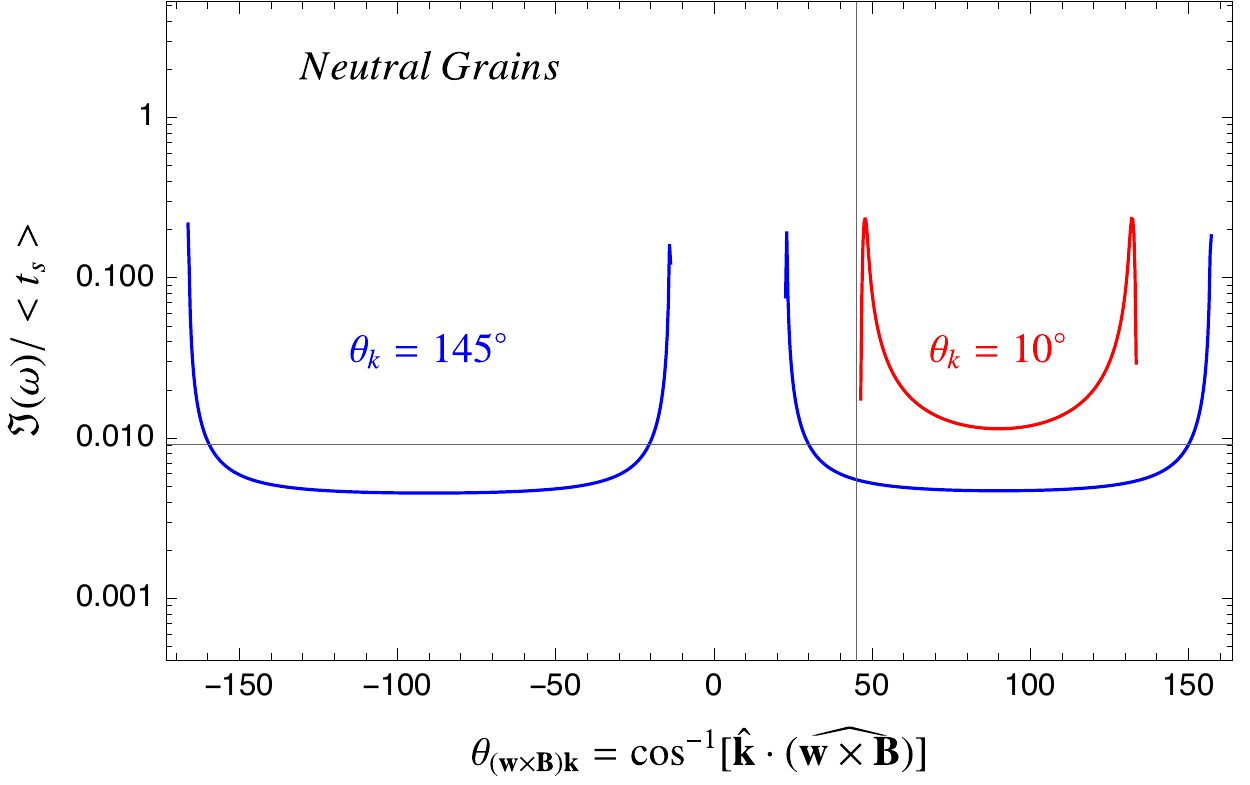}
\end{centering}\vspace{-0.3cm}
\caption{Linear-theory growth rates for the systems here, for a range of mode angles $\theta$ (directions of the wave-vector ${\bf k}$). We compare our fiducial case from Fig.~\ref{fig:3d_images} ({\em top}) and a case with identical parameters ($\driftvelmag$, $\ts$, and $\mu$) but with zero grain charge ({\em bottom}). The rates shown assume box-scale modes ($k=k_{0}=2\pi/\Lbox$) with fixed $\cos{\theta_{k}}\equiv \hat{\bf k}\cdot \driftvelhat$ (lines as labeled), as a function of the second angle (between $\hat{\bf k}$ and the direction mutually perpendicular to $\hat{\bf B}$ and $\driftvelhat$). The fastest-growing modes at the box scale have $\hat{\bf k}$ approximately aligned or anti-aligned with $\driftvel$ ($\theta_{k}\sim 0^\circ$ or $\sim 180^\circ$), without strong dependence on the second angle $\theta_{(w\times B)k}$ -- this means wavefronts should align into ``sheets'' perpendicular to $\driftvel$, as we observe. Without dust charge, the typical growth rates are suppressed by factors $\sim 100$, and the fastest-growing modes are sharply-peaked at specific $\theta_{(w\times B)k}$, producing distinct morphology.}
\label{fig:growth_rates}
\end{figure}

\begin{figure*}
\begin{centering}
\includegraphics[width=1.99\columnwidth]{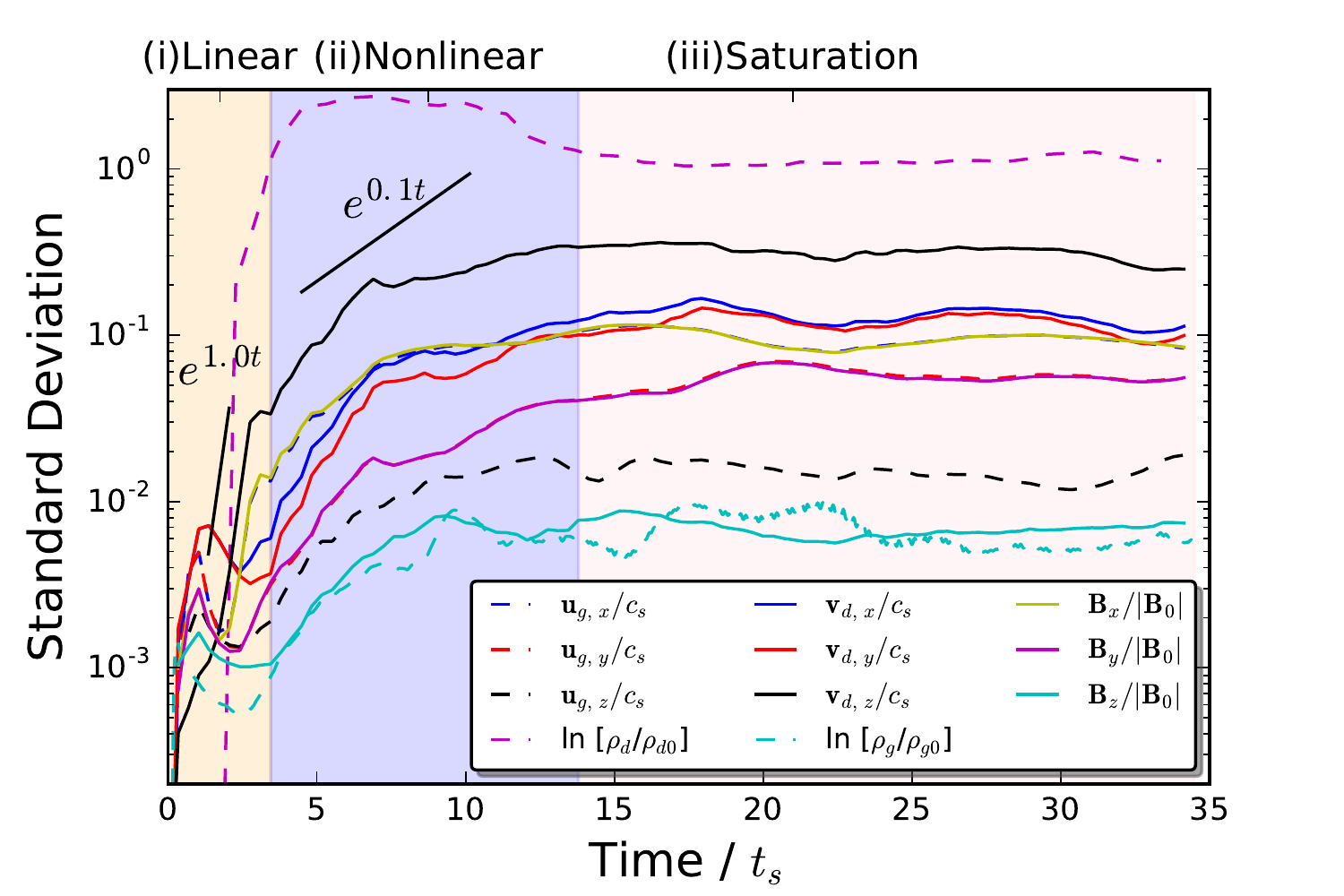}
\end{centering}\vspace{-0.3cm}
\caption{Volume-weighted rms standard deviation of various gas and dust properties vs.\ simulation time. We broadly denote three regimes in time: (i) linear, (ii) early nonlinear, and (iii) saturation. In (i), growth rates are rapid and agree reasonably well with the expectation from linear theory (Fig.~\ref{fig:growth_rates}) for a mix of modes from $k\sim (1-256)\,2\pi/\Lbox$ (we label $\sigma\propto e^{t}$ for comparison). In (ii) growth continues but at decreasing rates (see $e^{0.1\,t}$ for comparison), until saturating in (iii). There is strong anisotropy between e.g.\ $x/y/z$ fluctuations in $\gasvel$, $\dustvel$, $\B$. But for each component, the gas velocity and magnetic field fluctuations are tightly-coupled, especially perpendicular to $\B_{0}$. Dust velocities in the $\hat{\B}$ ($z$) direction fluctuate strongly, unlike gas. And while saturation is only weakly-compressible and sub-sonic in gas ($<1\%$ fluctuations in $\gasden$), it is highly nonlinear and compressible in dust ($\sigma[\ln{\dustden}] \gtrsim 1$).} 
\label{fig:varianceplot}
\end{figure*}

\begin{figure}
\begin{centering}
\includegraphics[width=0.99\columnwidth]{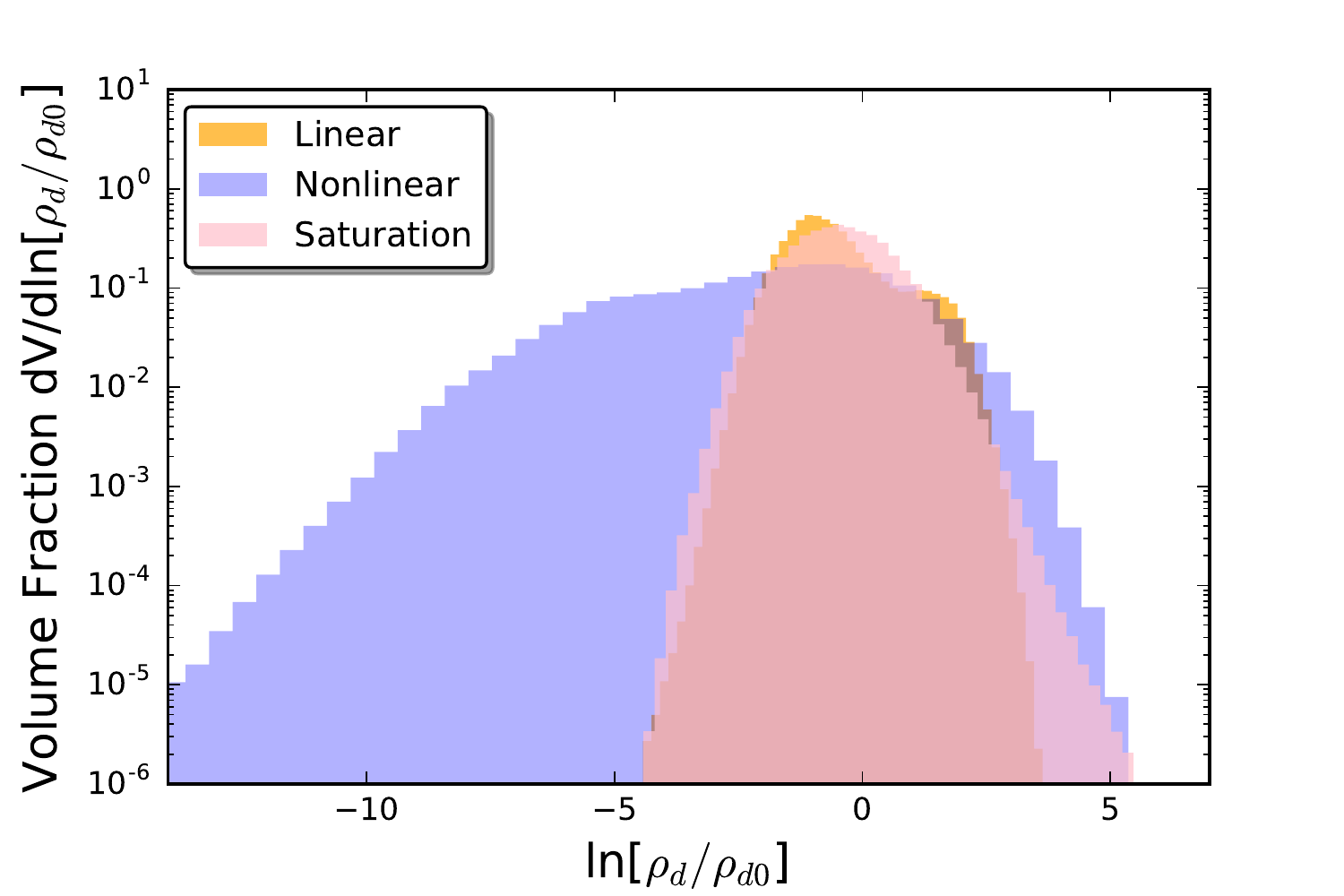}
\end{centering}\vspace{-0.3cm}
\caption{Volume-weighted probability distribution function (PDF) of dust density $\dustden$, evaluated halfway through each of the three time-regimes labeled in Fig.~\ref{fig:varianceplot}. Distributions are crudely log-normal, except in the tails. The large tail to low $\dustden$ in the early nonlinear stage occurs when the dust is maximally concentrated in ``sheets,'' leaving large dust-evacuated regions. Turbulence scatters some grains into these reducing the tail, but strong fluctuations persist.} 
\label{fig:pdf_plot}
\end{figure}

\subsection{Equations Solved}

We integrate the equations of motion for a population of charged grains in a magnetized gas. An individual grain satisfies,
\begin{align}\label{eq:eom.dust}
\Dt{\dustvel} &= \acc_{\rm ext,\,dust} + \acc_{\rm gas \mhyphen dust},
\end{align}
where $\Dt{}$ is the co-moving derivative, $\acc_{\rm ext,\, dust}$ is an external (constant) acceleration, and $\acc_{\rm gas\mhyphen dust}$ is the force from the gas on the grain. The latter is given by the sum of drag and Lorentz forces: 
\begin{align}\label{eq:drag.lorentz}
\acc_{\rm gas \mhyphen dust} &= -\frac{\driftvel}{\ts} - \frac{\driftvel\times\Bhat}{\tL},
\end{align}
where $\ts$ is the drag coefficient or \textit{stopping time}, $\tL$ the gyro or Larmor time, and $\driftvel \equiv \dustvel - \gasvel$ the \textit{drift velocity} (difference between grain velocity $\dustvel$ and gas velocity $\gasvel$). 

The gas obeys the ideal MHD equations, modified by the equal and opposite force from grains on gas (required by momentum conservation). In particular, the gas density $\gasden$ satisfies the usual advection equation, $\partial \gasden/\partial t = -\nabla \cdot( \gasden\,\gasvel)$, the magnetic field $\B$ satisfies the induction equation $\partial \B/\partial t = \nabla \times (\gasvel \times \B)$, and the momentum equation for $\gasvel$ is,
\begin{align}\label{eq:gas.mhd}
\nonumber \gasden \left( \frac{\partial}{\partial t} + \gasvel\cdot\nabla \right)  \gasvel =&\, -{\nabla P} - \frac{\B \times (\nabla \times \B) }{4\pi} + \gasden\,\acc_{\rm ext,\,gas} \\
 &- \int d^{3}\dustvel\,f_{d}(\dustvel)\,\acc_{\rm gas \mhyphen dust}(\dustvel,\,...).
\end{align}
Here $P$ is the gas (thermal) pressure, $\acc_{\rm ext,\, gas}$ is a (constant) external acceleration of the gas (we set this to zero in our simulations here), and the final term is the backreaction force on the gas from the grains, integrated over all grains at a given position. Here $f_{d}({\bf x},\,\dustvel)$ is the phase-space density distribution of dust, i.e.\ differential mass of grains per element $d^{3}{\bf x}\,d^{3}\dustvel$. The volumetric mass density of dust grains at a given position ${\bf x}$ is $\dustden \equiv \int d^{3}\dustvel f_{d}(\dustvel)$ (so as expected $\gasden\,\acc_{\rm dust \mhyphen  gas} = -\dustden\,\langle \acc_{\rm gas \mhyphen dust} \rangle$, where $\langle \acc_{\rm gas \mhyphen dust} \rangle$ is the mass-weighted average over all grains at that position). We assume an isothermal equation of state,  as this is usually a good approximation in most regions of interest (ISM, CGM, HII regions, etc.). 

In this simulation we assume Epstein drag (neglecting the Coulomb-drag contribution; see below). This can be approximated to very high accuracy with the expression (valid for both sub and super-sonic drift): 
\begin{align} \label{eq:ts}
\ts &\equiv \sqrt[]{\frac{\pi \gamma}{8}}\frac{\internaldensity \,\grainsize}{\gasden\,\cs}\, \bigg( 1+\frac{9\pi\gamma}{128} \frac{|\driftvel|^{2}}{\cs^{2}} \bigg)^{-1/2}.
\end{align}
Here $\internaldensity$ and $\grainsize$ are the \textit{internal} grain density and radius, respectively. The Larmor time is
\begin{align} \label{eq:tL} 
\tL &\equiv \frac{m_{\rm grain}\,c}{|q_{\rm grain}\,\B |} = \frac{4\pi\,\internaldensity\,\grainsize^{3}\,c}{3\,e\,|Z_{\rm grain}\,\B |},
\end{align}
where $m_{\rm grain}$ and $q_{\rm grain} = Z_{\rm grain}\,e$ are the grain mass and charge. In most regimes, the grain charge (for fixed composition and size, and a fixed radiation and/or cosmic ray background) depends primarily on the gas temperature (as compared to density, or velocity, or magnetic field; see \citealt{tielens:2005.book}). Since the gas here is isothermal, we therefore approximate $Z_{\rm grain}$ as constant.\footnote{Our fundamental assumptions (e.g.\ grains coupled via drag, ideal MHD) implicitly assume relatively large (i.e.\ non-PAH) grains in well-ionized environments, where $|Z_{\rm grain}| \gg 1$ is expected \citep{weingartner:2001.grain.charging.photoelectric}. This implies that the effects of charge ``flickering'' as a grain moves, important when $\langle |Z_{\rm grain}| \rangle \lesssim 1$, are not important.}

We  include only the Epstein drag contribution to dust dynamics, because in astrophysical contexts where one expects to see ionized plasma, the $0.1-1.0$ micron sized grains are safely out of the Stokes regime. We ignore the Coulomb drag, because in the supersonic drift regime, Epstein drag will dominate over the Coulomb drag, and in the subsonic drift regime, the two act in the exact same manner (and the normalization of the drag is arbitrary in our idealized setup). We ran numerous convergence tests, and verified that the different available hydrodynamic solvers and resolution in GIZMO did not change the results of the simulation significantly.

\vspace{-0.5cm}
\subsection{Numerical Methods}

We solve these equations using the multi-method code {\small GIZMO} \citep{Hopkins2014},\footnote{A public version of the code, including all methods used in this paper, is available at \href{http://www.tapir.caltech.edu/~phopkins/Site/GIZMO.html}{\url{http://www.tapir.caltech.edu/~phopkins/Site/GIZMO.html}}} using the second-order Lagrangian finite-volume ``meshless finite volume'' (MFV) method for the gas (MHD), which has been well-tested on problems involving multi-fluid MHD instabilities, the MRI, shock-capturing, and more \citep{hopkins:mhd.gizmo,hopkins:cg.mhd.gizmo,hopkins:gizmo.diffusion,su:2016.weak.mhd.cond.visc.turbdiff.fx,su:fire.feedback.alters.magnetic.amplification.morphology}. We model dust using the usual ``super-particle'' method \citep[e.g.][]{carballido2008kinematics,johansen2009particle,bai2010effect, pan2011turbulent}, whereby the motion of each ``dust particle'' in the simulation follows Eq.~\ref{eq:eom.dust}, but each represents an ensemble of dust grains of size $\grainsize$ (in other words, we ``sample'' some finite, computationally feasible number of grains to explicitly integrate trajectories for). The numerical methods for this integration are described and tested in \citet{hopkins2016fundamentally,Lee2017,moseley:2018.acoustic.rdi.sims} and for Lorentz forces on grains, we adopt the usual Boris integrator. The ``back-reaction'' is straightforward: in a given timestep $\Delta t$, one solves the coupled dust-gas equation exactly for the momentum change $\Delta {\bf p}$ to a single ``super-particle'' grain moving through a (internally homogeneous) cell, then subtracts that momentum from the gas (like the usual hydrodynamic flux), guaranteeing manifest conservation.

In addition to code tests and validation in the references above, and resolution tests and comparison to analytic solutions below, we have re-run our simulation with varied numerical choices. This includes (1) a different hydrodynamic solver (the meshless-finite-mass or ``MFM'' method), (2) the variant ``constrained-gradient'' MHD scheme in \citealt{hopkins:cg.mhd.gizmo} for the MHD reconstruction, (3) using a naive explicit leapfrog integrator instead of the Boris integrator for the Lorentz forces, and (4) different initial conditions (glass-like instead of lattice initial particle configurations). None of these substantially alters our results.

\vspace{-0.5cm}
\subsection{Equilibrium Solution, Initial Conditions, \&\ Units}

In \citet{Hopkins2018}, we show that the equations solved here have equilibrium, homogeneous solutions with uniform gas density $\initvalupper{ \gasden } \equiv M_{\rm gas,\,box}/\Lbox^{3}$, dust density $\initvalupper{ \dustden } \equiv \mu \, \initvalupper{ \gasden }$ ($\mu$ is the dust-to-gas ratio), gas velocity $ \gasvel  = \initvalupper{ \gasvel } + \acc_{\rm ext,\,gas}\,t + \acc\,\mu\,t / (1 + \mu)$ (where $\initvallower{ X } \equiv \langle X(\initvalupper{ \gasden },\,\initvalupper{ \driftvel },\,...,\,t=0) \rangle$ is the initial homogeneous value of $X$, and $\acc \equiv \acc_{\rm ext,\,dust}-\acc_{\rm ext,\,gas}$), and dust drift:
\begin{align}\label{eq:ws}
\initvalupper{ \driftvel } = \frac{\acc\,\initvalupper{ \ts }}{1+\mu} \, \left[ 
\frac{\hat{\acc} - \tau\,(\hat{\acc} \times \initvallower{ \Bhat }) + \tau^{2}\,(\hat{\acc}\cdot \initvallower{ \Bhat })\, \initvallower{ \Bhat } }{1+\tau^{2}}
\right]
\end{align}
where $\tau \equiv \initvalupper{ \ts } / \initvalupper{ \tL }$. 

Our simulations begin from these equilibrium solutions at $t=0$: we initialize a 3D periodic (cubic) box of side-length $\Lbox$ with uniform dust and gas densities, $\initvalupper{\gasvel} = 0$, and $\driftvel = \initvalupper{ \driftvel }$. Our  simulation uses $N_{\rm gas} = 256^{3}$ resolution elements for gas and an equal number for dust. For the sake of simplicity, and to facilitate physical understanding, the grains in the simulation are uniform in size and charge. 

We can make the equations solved dimensionless by working in units of the equilibrium sound speed $\initvalupper{ \cs }$, gas density $\initvalupper{ \gasden }$, and box length $\Lbox$. Then, for a given equation-of-state, it is straightforward to see that the dynamics of the problem (at infinite resolution) are entirely determined by six dimensionless constants: (1) the acceleration $\acceldl \equiv |\acc |\,\Lbox / (\initvalupper{ \cs })^{2}$, (2) the grain surface density or ``size parameter'' $\grainsizedl \equiv \internaldensity\,\grainsize / \initvalupper{ \gasden }\, \Lbox$, (3) the grain ``charge parameter'' $\grainchargedl \equiv 3\,\initvalupper{ Z_{\rm grain} } \,e / (4\pi\,c\,\grainsize^{2}\,(\initvalupper{ \gasden })^{1/2})$, (4) the dust-to-gas ratio $\mu \equiv \initvalupper{\dustden}/\initvalupper{\gasden}$, (5) the plasma $\beta \equiv \initvallower{P} / (|\initvallower{\B}|^{2}/8\pi)$, and (6) the angle $|\cos{\theta_{\B\acc}}| \equiv |\initvallower{\Bhat}\cdot\hat{\acc}|$ between the initial field direction $\initvallower{\Bhat}$ and $\hat{\acc}$. For the numerical simulation here, we use the input parameters\footnote{In our previous linear analysis, we defined $\beta\equiv c_{s}^{2}/v_{A}^{2}$ for convenience of notation, which is equal to unity in the simulation here, but differs from the more typical $\beta=P_{\rm thermal}/P_{\rm magnetic} = P_{0}/(|\initvallower{\B}|^{2}/8\pi)$ by a factor of $2$ for isothermal gas.} $\acceldl=5,\,\grainsizedl=5,\,\grainchargedl=50,\,\mu=0.01,\,\beta=2,\,\theta_{\B\acc}=87^\circ,\,\gamma=1$. Using Eq.~\eqref{eq:ws}, we can work out the  physical parameters of the equilibrium configuration as approximately $t_s=2.9\Lbox/\cs,\,\tau=30,|\driftvel|=0.9\cs,\theta_{\driftvel\B}=33^\circ$. This parameter set is mathematically equivalent to the input parameters but more convenient for linear-theory calculations.

These parameters are somewhat arbitrary but chosen for several reasons. (1) They are reasonable parameters for ``realistic'' $\sim 0.1$ micron grains in HII regions around luminous O-stars (see \citealt{Hopkins2018}). (2) The drift is (mildly) sub-sonic and sub-\Alf{ic}, so in the nonlinear regime we expect both sub and super-sonic movement of grains (which can produce distinct behaviors). (3) The equilibrium drift is sub-sonic, meaning the acoustic resonance does not exist but the \Alf{ic} but slow resonances do, while $\tau\gg 1$ implies that Lorentz forces strongly dominate drag forces.
As we show below, this causes  the behavior to differ radically from an un-charged or un-magnetized simulation. (4) With these parameters in linear theory, the slow magnetosonic and \Alf\ MHD-wave RDIs, the slow and \Alf\ gyro RDIs, the ``drift ($\driftvelhat$)-aligned''  (acoustic-like) and ``field ($\hat{\bf B})$-aligned'' (cosmic ray-like) modes are {\em all} present with comparable growth rates over the resolved dynamic range of the box, but have completely different mode eigenvectors and resonant angles/structures. This makes it an especially interesting case study, with a variety of mutually interacting modes and saturation mechanisms, and also makes prediction from linear theory especially difficult.

All statistics computed here are volume-weighted, and volumetric quantities for dust (e.g.\ the local dust density $\dustden({\bf x},\,t)$) are computed in post-processing from a local, adaptive kernel density estimator as described in \citet{moseley:2018.acoustic.rdi.sims}.

\begin{figure}
\begin{centering}
\includegraphics[width=0.99\columnwidth]{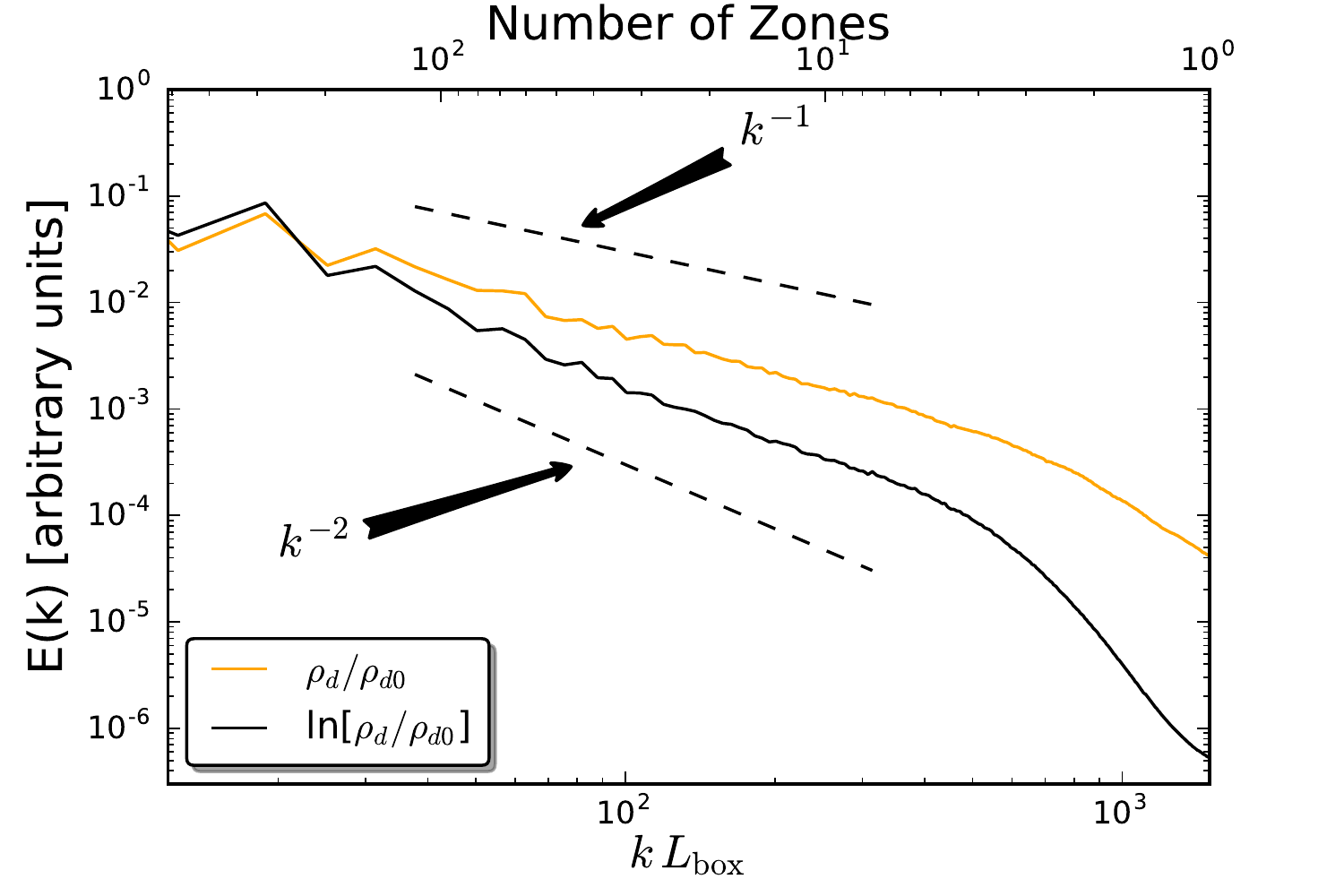}
\end{centering}\vspace{-0.3cm}
\caption{Power spectra ($E(k)$) of dust density ($\dustden$ and $\ln{(\dustden)}$), measured in the saturated state ($t>30\,t_{s}$) of our fiducial run. $E(k)$ is calculated after projecting simulation quantities on a $512^{3}$ Cartesian grid (there are ``spikes'' at high-$k$, which are artifacts of this projection, that we removed from the plot). We label wavenumber $k$ and equivalent number of zones/resolution elements in the gas per wavelength, and show power-law slopes for reference. As evident by-eye in Fig.~\ref{fig:3d_images}, although there is very fine structure within the dust ``sheets'' on small scales, the power is dominated by the largest (box-scale) modes.}
\label{fig:density_powerspectrum}
\end{figure}

\begin{figure}
\begin{centering}
\includegraphics[width=0.99\columnwidth]{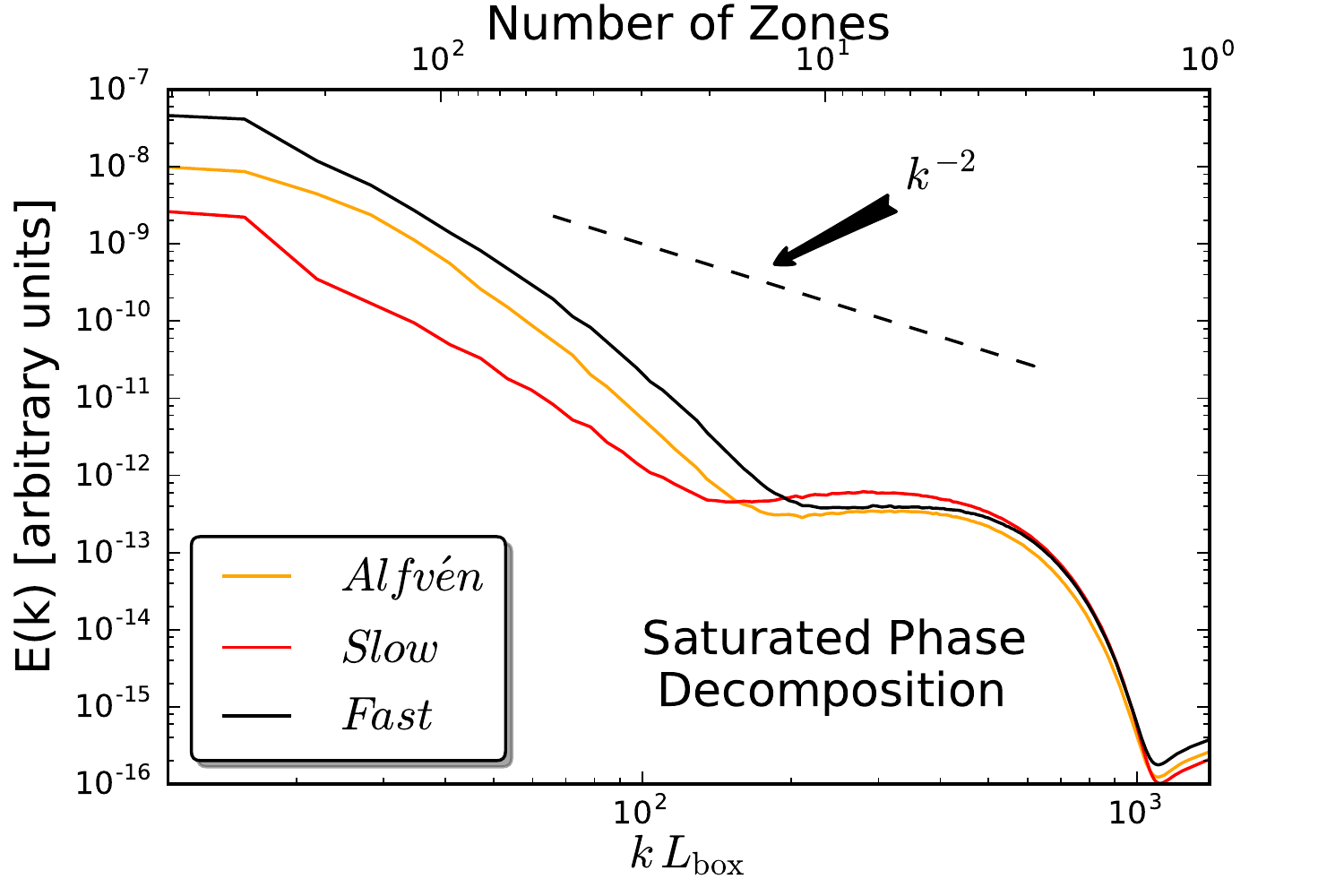}
\end{centering}\vspace{-0.3cm}
\caption{Power spectral density in \Alf{ic}, slow, and fast modes in gas in the  saturated state of the  fiducial run (as in Fig.~\ref{fig:density_powerspectrum}). This is obtained by projecting the velocity fluctuations in Fourier space (see \S~\ref{sec:analysis}). Power is  primarily on large scales, with a steep spectrum of all modes ($\sim\!k^{-2}$ or steeper) compared to standard low Mach number magnetized turbulence \citep{Schekochihin:2009eu}. The turbulence is dominated by a mix of \Alf{ic} and fast modes, with surprisingly weak excitation of slow modes (despite its sub-sonic nature).}
\label{fig:ModeDecomposition}
\end{figure}

\vspace{-0.5cm}
\section{Analysis}
\label{sec:analysis}

We present three-dimensional visualizations of the simulation of the magnetic RDI in Figure \ref{fig:3d_images}, at times corresponding to the (i) linear, (ii) early nonlinear, and (iii) saturation regimes.  The color scale at each projected slice corresponds to the strength of the magnetic field, and points indicate the location of the dust particles on the slice. Visualizations  of individual components of the magnetic field and velocity, as well as the density, are shown in Figure \ref{fig:2d_images}. These illustrate how the instability develops with a mix of Alfv\'enic and compressible modes. 
To emphasize the effect of the magnetic field on the nature of the instability, we provide an analogous purely hydrodynamical simulation in Figure \ref{fig:3d_images_hydro} at times corresponding to the (i) linear , (ii) early nonlinear, and (iii) saturation regimes. The initial conditions for this comparison simulation were chosen such that the drift velocity and strength of the turbulence would match that of the magnetized simulation. The  physical parameters of the equilibrium configuration are approximately $t_s=3.1\Lbox/\cs$ and $|\driftvel|=1.6\cs$. 

In Figure \ref{fig:growth_rates} we show the growth rates of the instability for our simulation initial conditions as predicted from the analytic linear stability theory. We show the growth rate for modes with wavenumbers  $k=k_0=2\pi/L$ and for a selection of mode angles $\theta_k=10,\, 55,\, 100$ and 155 for charged and uncharged  grains in the top and bottom panel respectively. Here, ${\bf k}$ is the mode wavevector, $\theta_k$ is  the cosine of the angle between vectors ${\bf B}$ and ${\bf k}$, and  $\theta_{(w\times B)k}$ is the cosine of the angle between vectors ${\bf w\times B}$ and ${\bf k}$. For a detailed description of the prescription to calculate these growth rates, we guide the reader to \citet{Hopkins2018}.

To quantify the different stages of the instability, we show the standard deviation of key physical parameters throughout the entire magnetized simulation in Figure \ref{fig:varianceplot}.  We show the temporal evolution of the three components of the magnetic field and velocity of the gas and dust, as well as the gas and dust density. The standard deviations generally evolve as $\sigma\sim e^{\omega_i t}$, with $\omega_{i}$ initially close to its linear-theory value but declining until saturation where growth ceases. In Figure \ref{fig:pdf_plot} we show the probability distribution function of (log) dust density during the three regimes in the simulation, which produces the most dramatic fluctuations during early nonlinear stages, before turbulence produces a more regular log-normal distribution.
Figure \ref{fig:density_powerspectrum} extends this by measuring the power spectrum of $\dustden$ and $\ln(\dustden)$,\footnote{We calculate the power spectrum as $E({|\bf k}|)=|\hat{\rho_d}({\bf k})|^2/(2\pi N^3)$, where ${\bf k} = \sqrt{k_x^2+k_y^2+k_z^2}$ and $k=2\pi/\lambda$, after projecting the simulation properties onto a $512^{3}$ Cartesian mesh (using the usual kernel density estimator at each mesh point to determine the local dust density). We verified that the power spectrum shape is not particularly sensitive to the resolution of this ``projection mesh.''} which shows most of the power is on large scales (as expected, given the prominent ``sheets''). Figure~\ref{fig:ModeDecomposition} presents the power spectrum of gas velocity decomposed into \Alf{ic}, slow, and fast modes.\footnote{We project the Fourier transformed gas velocity onto the basis 
\begin{align}
    \hat{\boldsymbol{\xi}}_A &\propto {{\bf k}}_\perp \times {{\bf k}}_\parallel \\
    \hat{\boldsymbol{\xi}}_s &\propto (1+\beta\gamma/2-\sqrt{D})\, {{\bf k}}_\perp + (-1+\beta\gamma/2-\sqrt{D})\, {{\bf k}}_\parallel \\
    \hat{\boldsymbol{\xi}}_f &\propto (1+\beta\gamma/2+\sqrt{D})\, {{\bf k}}_\perp + (-1+\beta\gamma/2+\sqrt{D})\, {{\bf k}}_\parallel 
\end{align}
Here, $\hat{\boldsymbol{\xi}}_A$, $\hat{\boldsymbol{\xi}}_S$ and $\hat{\boldsymbol{\xi}}_f$ represent the \Alf ic, slow and fast basis respectively (normalized to unit vectors); ${{\bf k}}_\perp$ and ${{\bf k}}_\parallel$ are the components of ${\bf k}$ perpendicular and parallel to $\B_{0}$; and $D=(1+\beta\gamma/2)^2-2\beta\gamma\,|{\bf k}_{\parallel}|^{2}/|{\bf k}|^{2}$. A derivation of these is provided in Appendix~A of \citet{Cho2003}, and we verified the procedure in idealized single-mode tests. Note that this decomposition assumes that $|\delta\B| \ll |\B_{0}|$, but this is true in our simulation.}

\vspace{-0.5cm}
\section{Discussion}

The predicted linear-theory growth rates for box-scale modes (Figure \ref{fig:growth_rates}) provide a reasonable approximation to the growth in fluctuations in the linear phase of the simulation (Figure \ref{fig:varianceplot}), despite the fact that all wavelengths are unstable. Moreover, the coherent 2D dust sheets can be qualitatively understood from linear theory, as the fastest-growing linear modes on the box scale are the ``aligned'' magnetic and drift modes (modes with wavevectors approximately aligned or anti-aligned with $\B$ and $\driftvel$, respectively, which are similar because the dust is strongly coupled to the fields), which produce aligned wavefronts (sheets) perpendicular to the drift. At scales $\sim \Lbox/10$, the \Alf\ RDI becomes the fastest-growing mode, and its fastest-growing mode angle is nearly perpendicular to $\hat{\bf B}$ (this generically occurs when the drift is subsonic; see \citealt{Hopkins2018} for details) -- this produces the ``corrugations'' in the sheets seen in Fig.~\ref{fig:3d_images} (especially at earlier times). If we zoom even further into the fine structure of the dust, around $\sim \Lbox/200$ the ``gyro'' resonances become dominant and produce serrations of the dust mutually perpendicular to both larger-scale modes. 

As shown in Figure \ref{fig:3d_images},  the dust self-organizes into coherent two-dimensional ``sheets'' at the onset of the instability, and these structures persist well into the saturation regime. In Figure \ref{fig:density_powerspectrum}, we quantify this by computing a formal power spectrum and confirm that the linear and natural log of the dust density predominantly have power on large scales throughout the simulation (likewise for gas, driven by dust; Figure \ref{fig:ModeDecomposition}). The power spectral linear and log density appears to decline  as $\sim k^{-1}$ and  $\sim k^{-1}\rightarrow k^{-2}$ respectively. At this stage, however, it is unclear if we have converged on an inertial range, and we plan to perform further higher-resolution studies in the future. We see in Figure \ref{fig:pdf_plot} that the distribution of the dust density develops significant non-Gaussian  tails in the early nonlinear and saturation phases. The largest discrepancies between these two phases manifest in the low density tail of the distribution, while the high density tail does not drastically evolve. This may be understood physically, since the instability tends to separate the dust into distinct sheets. During the early nonlinear phase, the instability is maximally efficient because the gas is not fully turbulent, and there are arbitrarily low dust to gas ratios between the sheets. During the saturation regime, however, the gas has fully developed nonlinear turbulence, and random gas motions tend to fill the under-dense regions with a small amount of dust, effectively removing the low density tail of the distribution.

The nonlinear evolution of the magnetized case is very different than the analogous acoustic case.  The growth rates corresponding to the same initial conditions, but with the grain charge set to zero, are orders of magnitude lower (Figure \ref{fig:growth_rates}). If we remove magnetic fields entirely but set up an analogous hydrodynamic run designed to have the same equilibrium $t_{s}$ and $\driftvel$ (Figure \ref{fig:3d_images_hydro}), the acoustic RDI manifests eventually (after growing much more slowly), but the geometric structure of the dust is quite different, and the dust concentration is vastly weaker. Our fiducial simulation is also qualitatively distinct from cases examined in \citet{Lee2017}, which included externally-driven MHD turbulence with similar Mach number but without the appropriate back-reaction from the dust on the gas (what actually drives the instabilities here). In all cases in that study, the dust density was either essentially uniform in the box, or strongly correlated with gas density, and at the (very small) gas Mach numbers here, the dust density fluctuations were $\sim 1\%$-level. Finally, we note that the gas turbulence seems quite different to standard theories of subsonic MHD/Alfv\'enic turbulence \citep{Sridhar:1994kg,Goldreich:1995hq,Schekochihin:2009eu}, where the cascade is dominated by Alfv\'enically polarized motions. Instead, as seen in Figure~\ref{fig:ModeDecomposition}, there is near equipartition of fast-wave and Alfv\'enic motions in the saturated turbulence, with a very steep spectrum (i.e. motions dominated by the largest scales). The dominance of fast modes, despite the modest Mach number, is interesting and is presumably related to their continued driving by the `drift-aligned'' mode on large scales.

Because the saturated turbulence is weak, the gas magnetic and velocity fluctuations are approximately linear, giving the perpendicular $\delta\B_{x,\,y}/|\B_{0}| \approx \delta {\bf u}_{x,\,y}/v_{p}$ and parallel $\delta\B_{z}/|\B_{0}| \approx (\delta{\bf u}\cdot{\bf k}_{\perp})/|\omega|$ (where $v_{p}\equiv |\omega/k_{\|}|$). Since the dominant modes have phase velocities of the sound or \Alf\ speed (and $v_{A}=c_{s}$ here), and the fastest growing modes have ${\bf k} \sim {\bf k}_{\|}$, this explains why we find $\delta\B_{x,\,y}/|\B_{0}| \approx \delta {\bf u}_{x,\,y}/\cs$, while $\delta\B_{z}/|\B_{0}|$ is suppressed by a factor $\sim k_{\perp}/k_{\|}$ (Figure \ref{fig:varianceplot}). Interestingly, Figure \ref{fig:2d_images} shows the density fluctuations and compressible (longitudinal, i.e.\ ${\bf u}_{z}$) fluctuations closely trace $\B_{z}$; this is expected if gas pressure fluctuations are in approximate equiparition with magnetic pressure fluctuations (as in e.g.\ fast modes). This suggests the relations  $\delta \ln\rho_g=(|{\bf B_0}|^2/c_s^2\rho_{0g})(\delta B_z/|{\bf B_0}|) \approx (\delta B_z/|{\bf B_0}|)$, and, using $d\ln\gasden/dt=-\nabla\cdot\gasvel$ with ${\bf k}\sim {\bf k}_{\|}$, $\delta {\bf u}_{z} \sim |\omega/k|_{\rm fast}\,\delta\ln{\gasden} \sim \sqrt{2}\,\cs\,\delta\ln{\gasden}$ (see Figure \ref{fig:varianceplot}). 

It is clear from Figure \ref{fig:2d_images} that the driving of the gas occurs as follows: the dust aligns and condenses into sheets, which (having locally high dust-to-gas ratio) are differentially accelerated more strongly in the acceleration direction (nearly perpendicular to $\B$). The sheets ``slide,'' dragging gas (and field lines) along, generating non-zero $\B_{x,y}$. Magnetic tension therefore is the relevant limiting process, and we might expect saturation when the ``driving'' force from the dust on the gas ($\sim \dustden\,\driftvel/\ts \sim \mu\,\gasden\,{\bf a}$) is balanced by tension forces ($\sim \B\times(\nabla\times\delta\B)/4\pi \sim k\,|\B_{0}|\,\delta\B/4\pi$) with box-scale modes dominant. Put together, this gives $\delta\B/|\B_{0}| \sim 4\pi\,\langle \dustden\rangle\,|{\bf a}|\,\Lbox/|\B_{0}|^{2} \sim \mu\,\acceldl\,\beta \sim 0.1$ (compare Figure \ref{fig:varianceplot}). 
This argument is only very approximate and likely an incomplete description, but it does appear that the simulations here saturate differently from the pure-hydrodynamic acoustic RDIs studied in \citet{moseley:2018.acoustic.rdi.sims}. There, the saturation occurred when box-scale eddy turnover times were approximately equal to mode growth times $\Im(\omega[k\sim 2\pi/\Lbox]) \sim 1/t_{\rm eddy} \sim \delta\gasvel/\Lbox$, which  would imply $\delta\gasvel \sim \cs$, an  order-of-magnitude larger than seen here. In other words, magnetic forces appear to limit the turbulence before it becomes so vigorous.

We do not, at present, have a predictive model for the saturated dust density fluctuations, although since the dust is collisionless, it is not surprising that it  clumps much more strongly than gas. In future work, we will explore analytic models for the saturated dust density and velocity structures.

Also in future work, we will explore changes to the gas thermodynamics (e.g.\ equation of state) and dust drag/charge laws. However, since in this particular case the gas density fluctuations are very small, changes in the equation of state should not have large effects. Moreover since temperature is the dominant local gas property that influences the grain charge, the charge would not vary strongly in this particular case if we adopted more complicated expressions for charge scalings. We also expect that varying the dust-to-gas ratio $\mu$  will not change the qualitative behavior of the instability (modulo slower growth rates and some shift in characteristic wavelengths), but the saturated behavior may be different, so this also merits exploration. However, linear theory suggests that qualitatively different behavior might emerge if we change parameters like the drift velocity and ratio of Lorentz-to-drag forces ($\tau^{0}\sim 30$, here): these determine which modes are the fastest-growing, as well as their characteristic fastest-growing angles and eigenstructure. The parameters here are plausible for some astrophysical regimes including parts of HII regions, the CGM, and supernovae remnants at specific times in their expansion, but an enormous diversity of these parameters exists astrophysically (with e.g.\ $\tau \sim 10^{-10} - 10^{10}$ plausible in different dusty astrophysical environments; see \citealt{Hopkins2018}). In future work, therefore, it will be extremely interesting to further explore this broad parameter space.

\vspace{-0.5cm}
\acknowledgments{This work was initiated as part of the Kavli Summer Program in Astrophysics, hosted at the Center for Computational Astrophysics at the Flatiron Institute in New York. We thank the Kavli Foundation and the Simons Foundation, for their support. DS thanks  Fred Adams, Andrea Ferrara and Daniel Lecoanet for insightful comments and suggestions that significantly contributed to this work. Support for PFH was provided by an Alfred P. Sloan Research Fellowship, NSF Collaborative Research Grant \#1715847 and CAREER grant \#1455342, and NASA grants NNX15AT06G, JPL 1589742, 17-ATP17-0214. Numerical calculations were run on the Caltech compute cluster ``Wheeler,'' allocations from XSEDE TG-AST130039 and PRAC NSF.1713353 supported by the NSF, and NASA HEC SMD-16-7592.}




\bibliographystyle{mnras}
\bibliography{refs} 





\bsp	
\label{lastpage}
\end{document}